
\input phyzzx
\doublespace
 \newtoks\slashfraction
 \slashfraction={.13}
\def\lm{\lambda}
\def\kp{\kappa}
\def\gm{\gamma}
\def\om{\omega}
\def\ll{\lbrack}
\def\rr{\rbrack}
\def\prt{\partial}
\def\prtp{\prt_+}
\def\prtm{\prt_-}
\def\prtpm{\prt_\pm}
\def\prtmp{\prt_\mp}
\def\nab{\prt^2}

\def\prm{\prime}
\def\upm{^\prime}
\def\al{\alpha}
\def\be{\beta}
\def\eps{\epsilon}
\def\xos{{{\prt x}\over{\prt\sigma}}}
\def\xot{{{\prt x}\over{\prt\tau}}}
\def\hxos{{{\prt\hat x}\over{\prt\sigma}}}
\def\hxot{{{\prt\hat x}\over{\prt\tau}}}
\def\xost{{{\prt^2 x}\over{\prt\sigma\prt\tau}}}
\def\hxost{{{\prt^2\hat x}\over{\prt\sigma\prt\tau}}}

 \def\slash#1{\setbox0\hbox{$ #1 $}
 \setbox0\hbox to \the\slashfraction\wd0{\hss \box0}/\box0 }
\pubnum={5647}
\date{August 1991}
\titlepage
\title{Self-Gravitating Strings \break In $2+1$ Dimensions
 \foot{Work supported by the Department of Energy, contract
 DE-AC03-76SF00515.}}
\author{Shahar Ben-Menahem}
\medskip
\address{Theory Group
\break Stanford Linear Accelerator Center
\break P.O.B. 4349, Stanford University
\break Stanford, CA 94309}
\medskip
\abstract
 We present a family of classical spacetimes in $2+1$ dimensions.
 Such a spacetime is produced by
  a Nambu-Goto self-gravitating string. Due to the special properties
  of three-dimensional gravity, the metric is completely described
  as a Minkowski space with {\it two} identified worldsheets.
   In the flat limit, the standard string is recovered.
  The formalism is developed for an open string with massive endpoints,
   but applies to other boundary conditions as well.
    We consider another limit,
 where the string tension vanishes in geometrical units but the
  end-masses produce finite deficit angles. In this limit, our open
 string reduces to the
 free-masses solution of Gott, which possesses closed timelike curves
 when the relative motion of the two masses is sufficiently rapid.
 We discuss the possible causal structures of our spacetimes in
 other regimes.
  It is shown that the induced worldsheet Liouville mode
  obeys ({\it classically}) a differential equation, similar to
  the Liouville equation and reducing to it in the flat limit.
  A quadratic action formulation of this system is presented.
 The possibility and significance of quantizing the self-gravitating
 string, is discussed.
\endpage

\REFS\sh{For some references on discrete and continuum works in
  $2d$ gravity, see S. Ben-Menahem, SLAC-PUB-5262(1990), to appear
 in Nuclear Physics.}\refsend
\REFS\lnrs{A.R. Cooper, L. Susskind and L. Thorlacius, SLAC-PUB-5536,
 and references therein.}\REFSCON\Polch{J. Polchinski,\journal
 Nucl.Phys.&B324(1989)123.}\refsend

\REFS\Deser{S. Deser, R. Jackiw and G.`t Hooft,\journal Ann.Phys.
&152(1984)220.}\refsend
\REFS\Gott{J.R. Gott,\journal Phys.Rev.Lett.&66(1991)1126.}\refsend
\REFS\wtn{E. Witten,\journal Nucl.Phys.&B311(1988)46; \journal
 Nucl.Phys.&B323(1989)113.}\refsend
\REFS\thred{A.A. Migdal and M.E. Agishtein,\journal Mod.Phys.Lett
&A6(1991)1863.}\refsend
\REFS\Clement{G. Clement,\journal Ann.Phys.&201(1990)241 and
 references therein.}\refsend
\REFS\Hawking{S.W. Hawking and
 G.F.R. Ellis, `The Large Scale Structure
  of Spacetime', Cambridge University Press, 1973.}\refsend
\REFS\Teitel{T. Regge and C. Teitelboim,\journal Ann.Phys.&88
(1974)286, and references therein.}\refsend
\REF\Henn{J.D. Brown and M. Henneaux,\journal Comm.Math.Phys.&104
(1986)207.}\refsend
\REFS\Thorne{M.S. Morris, K.S. Thorne and U. Yurtsever,\journal
Phys.Rev.Lett.&61(1988)1446.}\refsend
\REFS\Carter{B.Carter,\journal Phys.Rev.&174(1968)1559.}\refsend
\REFS\Friedman{J.L. Friedman {\it et al},\journal Phys.Rev.&D42(1990)
1915.}\refsend
\REFS\Weinberg{S. Weinberg, `Gravitation and Cosmology',Wiley, N.y.
 1972.}\refsend
\endpage

\chapter{Introduction}
Much attention has been devoted in recent years to the study of
low dimensional gravity ( two or three spacetime dimensions). In two
 dimensions, gravitation is classically trivial, but the quantum
  conformal anomaly renders the Liouville mode dynamical,
  and the structure of the theory (with various forms of matter)
  is both rich and tractable $\ll 1\rr$.
   Two-dimensional
 gravity is useful both as a toy model to guide our thoughts
 about the (as yet nonexistent) four-dimensional theory $\ll 2\rr\ll
3\rr$,
  and as a new
 tool for studying noncritical string theory and string field theory.
\par
  In $2+1$-dimensional Einstein gravity,
   there is still no dynamical graviton, but the
 classical theory is already nontrivial: a nonvanishing
  energy-momentum tensor may exist\foot{Even with no cosmological
 constant.},
  and can be remotely detected by parallel-transporting
     tensor quantities around  the matter distribution. Another
globally-nontrivial aspect of spacetime, when moving sources are present,
 is an obstruction to defining global Minkowskian time, which results in
 `time shifts' experienced by observers (external or part
  of the system)
 travelling along some closed paths $\ll 4\rr$. Some matter
  distributions, although satisfying the energy-positivity conditions,
 give rise to sufficient
 time shifts to allow `naked' closed timelike or lightlike
  curves$\ll 5\rr$.
 In addition, a topological mass-term is allowed in three dimensions,
 which {\it does} render the graviton dynamical.
 \par Classical $2+1$ dimensional gravity is also relevant as an
  approximation scheme (dimensional reduction) in the study of
 cosmic strings; an idealized infinite, straight cosmic string is a
  three-dimensional pointlike particle.

  {\it Quantum} Einstein gravity has also been
   studied in three dimensions,
   both in the continuum$\ll 6\rr$ and discretized$\ll 7\rr$
   versions.
\par
 In this paper, we initiate the study of
  classical three-dimensional spacetimes
 produced by a particular type of matter --- a single, Nambu-Goto
 string. In other words, the action is the Einstein-Hilbert term
 plus the Nambu-Goto action, the latter being covariantized in the
 three target-space dimensions.
  We shall refer to this as the self-gravitating, or
   Nambu-Goto-Einstein, string.
     There are several reasons for considering this system. It has
 a richer dynamics than the classical matter distributions thus far
 considered in three-dimensional gravity$\ll 4\rr\ll 8\rr$ .
  In addition, if an {\it open} string with two massive
 endpoints is considered, one recovers in the zero-tension limit, the
 spacetime recently considered by Gott$\ll 5\rr$.
  This spacetime simply consists
  of two point-masses (conical singularities)
  moving past each other. When the relative motion is
 sufficiently rapid, the spacetime possesses closed timelike curves.
 It is therefore of interest to study the causal structure of
 spacetime in the case where the masses are tethered together elastically
 --- since then their relative motion is nonuniform. The mathematically
 simplest way to implement this generalization, seems to be via the
 Nambu-Goto-Einstein string.
  \par
   As a final motivation, we
   wish to raise the question of whether the self-gravitating
 string can be consistently quantized\foot{Since this is
 a 3d string theory, we might well have to add a CFT on the
 worldsheet as a necessary condition for the theory to be consistent.}
  --- and if so, whether it
 is relevant to ordinary string theory.
 \par
   The main results reported here are as follows.
    The geometry of our classical
     spacetime is expressed as a flat Minkowski
 space, with the region between two worldsheets excised and the two
 worldsheets identified. The equations of motion
  of the worldsheet
 are found, and are reduced to a Liouville-like equation for the
 induced worldsheet Liouville mode.
  The flat limit and the zero-tension limit are worked out
 for the case of open string with massive endpoints.
 It is shown how to expand solutions in the geometrized string tension.
  The two-worldsheet formalism is recast, using auxiliary fields,
   as an action principle on the worldsheet. The new action is
  quadratic, with quadratic constraints on the fields,
   and is perhaps amenable to consistent quantization.
  It is suggested that the {\it first}
   quantized self-gravitating string could be
 a step towards understanding nonperturbative string field theory in
 a continuum setting, if such a quantization is possible.
\par
    Further work concerning the possible global structures
  (causal and topological) of
   our classical spacetimes, and the feasibility of
 the proposed quantization scheme, is in progress.
\par
 Throughout, we concentrate mostly on the case of open string with
 massive endpoints, because our original motivation was in extending the
 Gott solution. In the context of quantization, of course, one should
 work either with standard open strings (that is, having lightlike
 endpoints), or with closed strings; The treatment differs only in the
 boundary conditions, and we indicate the requisite changes at the
 appropriate places.
 \par
  The remainder of this paper is organized as follows. In section 2,
 the geometry of our spacetime is described, and the attendant
  formalism set up. In section 3 the string equations of motion
 are derived, and the geometric meaning of the masses at the ends
 of the open string is established.
  In section 4 we state how the equations of motion can be reduced
 to a ({\it classical}) Liouville-like equation; the details of this
 demonstration are relegated to appendix B.
 Section 5 contains a discussion of the asymptotic form of
  the three-metric in 2+1 dimensional gravity, and of the global
 structure of spacetime.
  In section 6 we treat
 perturbation theory in the string tension, and derive the flat limit
 and the zero-tension limit for
 the case of open string with massive endpoints.
 In section 7, the possible causal structures
  of our spacetimes are discussed, as well as
  the general significance of a `nonphysical'
 causal structure.
 In section 8, the quadratic-action formulation is presented; some
 off-shell results from appendix B are used. In section 9
 we discuss quantization, and section 10 summarizes the conclusions.
 Appendix A lists our Minkowski-space, curvature and
 units conventions. In appendix C we present a proof that the
 geometrically-defined endpoint masses are conserved, whereas
  appendix D treats the flat open string with endmasses, and presents
 a concrete configuration of such a string.
\chapter{Geometry of Self-Gravitating String.}
 As explained in the introduction,
 we shall mostly treat the case of open string with massive endpoints,
 but will indicate how the more familiar cases
  (open string with lightlike
 ends, and closed string) are to be handled. The only differences are
  in the boundary conditions imposed,
 and in the global properties of the resulting spacetime\foot{Since
 the open string with massive endpoints is not a familiar case, even in
 the flat limit, the reader is referred to appendix D for a brief
 treatment of the flat case for such a string.}.
\par
The spacetime we are interested in has the topology of $R^3$;
its metric has
 Minkowskian signature, and is everywhere flat except on a worldsheet
 of an open string. Fig. 1 depicts the string, which is an equal-time
  section of
 the worldsheet. We denote its two endpoints $P$ and $Q$.
  We also use this
 notation for the {\it worldlines} of the respective endpoints, which
constitute
 the boundary of the worldsheet (the string is assumed to exist in
 the infinitely remote past or future)\foot{Since we are interested in
 possible closed timelike curves, caution should be exercised in using
 terms such as `section', `past' or `future'. Since spacetime is assumed
 homeomorphic to $R^3$, a global coordinate system can be chosen.
  A global
 time coordinate does not always exist, but we assume that it {\it does}
 exist in some open neighborhood of the worldsheet itself. Thus it makes
 sense, in our spacetimes, to talk about equal-time sections of the
 worldsheet, but not necessarily of the entire manifold; it is in this
 sense that Fig.1 is to be understood. Of course, there is in general
  no global killing vector, temporal or other, so the equal-time
 sections of the worldsheet depend on the time coordinate chosen.}.
 We denote the worldsheet by $S$; it is assumed to have the topology
 $I\times R$, where $I$ is the closed unit interval\foot{We choose to
disallow self-intersecting worldsheets. The reason is that if the
string cuts itself, a new curvature singularity is created at the
intersection, and in addition the global structure of spacetime
changes. This is in marked contrast with standard string theory,
where self-intersections of the embedding of $S$ are irrelevant.
Therefore, {\it some} of the configurations resulting from our
construction may not be physical solutions; conversely, the
self-intersecting solutions will not be globally reproduced by our
construction. We will further discuss these global considerations in
section 5.}. For simplicity, we remove from spacetimes infinitesimal
neighborhoods of the singularities $P$ and $Q$.
\par
 The energy-momentum tensor is nonzero only on the worldsheet, so due
 to Einstein's field equations and the dimensionality of spacetime,
 space is locally flat everywhere else. We therefore choose the following
 convenient coordinatization of the manifold. Let an orientation be
 defined on the manifold; This induces an orientation on any
 equal-time section of a neigborhood $V$ of the worldsheet (see the
 previous footnote). The worldsheet is two-sided, and we define its
 {\it bottom} side as that through which any directed curve, lying
 within the section and winding
 around $S$ in a positive sense, passes on the segment from $P$ to
 $Q$.
 The opposite side will be referred to as
  the {\it top} of the worldsheet.
 Now, we erect a flat Minkowskian coordinate system on the bottom side
 of $S$, and extend it outside of $V$ and
  around $P$ and $Q$. However, the coordinates
 $\{z^\mu\}$ assigned in this way to a point on the top side, will depend
 on whether one continues around $P$ or $Q$. Therefore, transition
 functions are needed on the top side of $S$\foot{And also in some
 region outside $V$.}. In fact, the transition region can be chosen to
 intersect $V$
 anywhere on the top side, but must run from $S$ to infinity. Unlike $S$,
 it does not correspond to any physical singularity,
  so both $g_{\mu\nu}$ and $\Gamma^\lm_{\mu\nu}$ are continuous across
 it. Thus, the transition functions must constitute
 an isometry of the Minkowski
 metric, that is, a Lorentz transformation combined with a
 shift$\ll 4\rr$.
   We call this coordinate atlas $M$. It describes
the manifold with $S$ itself removed; the {\it entire} manifold is
$M$ with a homeomorphism that identifies the top and bottom sides of
 $S$.
 \par
 In $M$, the coordinates of the worldsheet differ on the top and
 bottom sides, whether one continues around $P$ or $Q$; we adopt
 the continuation around $P$.
  Let the bottom coordinates be $x^\mu=x^
\mu(u)$, where $u^a$ is a parametrization of the worldsheet. We
 denote by $\hat x^\mu(u)$ the coordinates of the point, on the top
 of $S$, that is to be identified with $x^\mu(u)$. Clearly, at $P$
 the two sets of coordinates must agree; thus one boundary condition is
 $$x^\mu(P)=\hat x^\mu(P)\eqno(1)$$
  The Minkowski space, with the appropriate identifications,
   is depicted in Fig. 2; the transition region, $T$, has been
  chosen to run from $Q$ to
 infinity, along an arbitrary continuation of the worldsheet. If the
 surface $T$ is considered as part of $S$, then
  for a
 point $u^a$ on the surface $T$ or its boundary $Q$, $x$ and $\hat x$
 are related by the linear transition function:
 $$x(u)=\bar L \hat x(u)+\bar b\quad on\; T\; ,\eqno(2)$$
 where $\bar L$ is a fixed Lorentz transformation, and $\bar b$ is a
 fixed three-vector. At $Q$, eq.(2) becomes the second
  boundary condition:
 $$x(Q)=\bar L \hat x(Q)+\bar b\eqno(3)$$

 We will occasionally refer to $x$,$\hat x$ as the two worldsheets ---
 bottom and top, respectively;
 although they describe the same surface.
  Clearly, we could just as well have chosen the transition
   region from $P$ to infinity, and continued around $Q$---
    in which case the boundary
  condition at $Q$ would have been the simpler one.
\par Following standard string practice,
 we choose the parametrization $u^a$ to be orthonormal on
 the bottom worldsheet; then, by the continuity of the
  three-dimensional metric
 across $S$, $u^a$ are orthonormal on the top worldsheet as well
\foot{$g_{\al\be}(z)$ are not continuous functions in the $\{z\}$
coordinates. To obtain continuous components, we need to continue
$\{z\}$ {\it across} $S$, rather than around $P$. We shall define
{\it two} such coordinate systems, below eq.(8d).}.
 We denote:
 $$ u^0=\tau,\; u^1=\sigma,\; u^\pm=u^0\pm u^1 \eqno(4.a)$$
 $$ \prt_\pm\equiv\prt/\prt u^\pm  \eqno(4.b)$$
 The orthonormality condition then reads (see appendix A for
 conventions),
 $$ (\prtpm x)^2=(\prtpm\hat x)^2=0\; . \eqno(5)$$
 Eq.(5) incorporates the continuity of the
  $++$ and $--$ components of the 3d
 metric at $S$; the continuity of the $+-$ component implies,
 $$ E\equiv\prtp x\cdot\prtm x=\prtp\hat x\cdot\prtm\hat x
 \eqno(6) $$ where
 $2E$ is the induced conformal scale factor on $S$: the induced metric
 is $ds^2=2Edu^+du^-$. \par
 There is still a residual parametrization freedom, which we partially
 use up in
 choosing $P$,$Q$ to be $\sigma=0$,$\pi$, respectively\foot
{ There still remains the freedom to reparametrize via $u^+\rightarrow
 f(u^+)$,$u^-\rightarrow f(u^-)$, where $f(v)-v$ is periodic
  with period $2\pi$.}.
\par Recall that we have assumed the worldsheet has global equal-time
 sections. In terms of the $u^a$ coordinates, we require that the lines
 $\tau=\; const$ are spacelike, whereas $\sigma=\; const$ are timelike.
 Since $u^a$ are orthonormal, this is equivalent to
 $$E(u)<0\; .\eqno(7)$$
 \par It is useful to define the normal to $S$. We define $n(u)$ as
 the unit (spacelike) three-vector normal to $S$ on its bottom side,
 and pointing into the {\it top} side, at point $u$. Since
   $\prtpm x$ span the local tangent plane to $S$, we find
(components in $\{z\}$ coordinates)
$$ n_\mu={1\over E}\eps_{\mu\al\be}\prtp x^\al \prtm x^\be
 \; ,\eqno(8a)$$
$$ n^2=1,\quad n\cdot\prtpm x=0 \; .          \eqno(8b)$$
 Similarly, the unit normal vector to the top side, also pointing
into the top side, is in the same coordinates (by virtue of eq. (6))
$$\hat n_\mu={1\over E}\eps_{\mu\al\be}\prtp \hat x^\al
 \prtm \hat x^\be\; ,\eqno(8c)$$
$$ \hat n^2=1,\quad \hat n\cdot\prtpm \hat x=0\; . \eqno(8d)$$
 In order to study the three-geometry in the neighborhood $V$ of the
 worldsheet, we extend the bottom-side coordinates, $\{z\}$,
 across $S$ into the top side, and call the new coordinate system
 $\{x\}$. Another useful coordinate system, $\{\hat x\}$, is obtained
 by extending $\{z\}$ from the {\it top} side across $S$.
  This notation extends in a consistent way our use of $\hat x,x$
 to denote the top and bottom coordinates of the surface $S$ itself,
 into $V$. Thus, we have $\hat x(x(u))=\hat x(u)$.

  Below $S$, $g_{\mu\nu}(x)=\eta_{\mu\nu}$, whereas above $S$,
  $\hat g_{\mu\nu}(\hat x)=\eta_{\mu\nu}$.
   But the metric $g$, in the $x$
 coordinate system, must be continuous at $S$. The continuity of
 the tangential components was already encoded in eqs.(5),(6)
 above; that of the normal-normal and mixed components, reads
 $$({{\prt \hat x}\over{\prt n}})^2=1, \eqno(9a)$$
 $${{\prt \hat x}\over{\prt n}}\cdot\prtpm\hat x=0 \eqno(9b)$$
 where we denote $\prt /\prt n\equiv n^\al\prt /\prt x^\al$.
   From eqs.(8d),(9a) and (9b) we easily find:
$${{\prt\hat x}\over{\prt n}}=\hat n. \eqno(9c)$$
  \par The connection, in $x$ coordinates, is
 $$\Gamma^\mu_{\al\be}={1\over 2}g^{\mu\lm}(g_{\al\lm,\be}+
 g_{\be\lm,\al}-g_{\al\be,\lm})\; .\eqno(10)$$
 The curvature on $S$ causes a discontinuity in $\Gamma$. Since the
 worldsheet is that of a Nambu-Goto string, and since $g_{\mu\nu}$ is
 Minkowskian at $S$, the energy-momentum tensor in the bulk of the
 worldsheet (that is, away from $P$,$Q$) is given in $\{x\}$
 coordinates by
 $$ T^{\mu\nu}(y)=\kp\int du^+du^-\bigl(\prtp x^\mu\prtm x^\nu+
 \prtm x^\mu\prtp x^\nu\bigr)\delta(y-x(u)), \eqno(11)$$
  where $\kp>0$ is the string tension\foot{This $T_{\mu\nu}$ satisfies
the weak, strong and dominant energy positivity conditions$\ll 9\rr$.
  }. We work in geometrized units, in
 which the three-dimensional Newton's constant is
 $$ 8\pi G=1,$$ masses are dimensionless, and
 $\kp$ has dimensions of inverse length. The field equations are thus
 (see appendix A)
  $$ R_{\mu\nu}-{1\over 2}g_{\mu\nu}R=-T_{\mu\nu}\eqno(12) $$
 Now, $g_{\al\be,\lm}$ vanishes on the bottom side of $S$; let its
  dicontinuity
   across $S$ (top value minus bottom value) be\foot{The
   discontinuity of the gradient of a function
  that is a constant on one side of $S$, is normal to the surface $S$.}
 $$\Delta g_{\al\be,\lm}\equiv n_\lm P_{\al\be}(u)\eqno(13a)$$
 Then by eq. (10),
 $$ \Delta\Gamma^\mu_{\al\be}={1\over 2}(n_\be P_\al^{\;\mu}+
 n_\al P_\be^{\;\mu}-n^\mu P_{\al\be}) \eqno(13b) $$
   Let $P^{(t)}$ denote the projection of $P$ on the plane locally
   tangent to $S$ at $u$. Eqs. (11)-(13) imply:
 $$P^{(t)}_{\mu\nu}=4\kp(n_\mu n_\nu-\eta_{\mu\nu})+
 {{2\kp}\over E}(\prtp x_\mu\prtm x_\nu+\prtm x_\mu\prtp x_\nu)
  \eqno(14)$$
\chapter{The String Equations of Motion.}
The equations of motion of the string are easily found by writing
the Nambu-Goto action in the background metric $g_{\mu\nu}$, and
varying this action w.r.t. $x$. We obtain, in $\{x\}$ coordinates,
$$\nab x^\mu+\bar\Gamma^\mu_{\al\be}\prtp x^\al\prtm x^\be=0,
  \eqno(15a)$$
 where $\bar\Gamma$ is half the discontinuity,
$$ \bar\Gamma^\mu_{\al\be}\equiv{1\over 2}\Delta\Gamma^\mu_{\al\be}.
\eqno(15b)$$
 $\bar\Gamma$ is also the average of the values of $\Gamma$ on the top
 and bottom sides. It is easy to verify that eqs.(15)
  ensure the local
 conservation of $T_{\mu\nu}$ in the bulk of the worldsheet.
\par
 We contract indices $\mu$,$\nu$ in eq.(14) with $\prtp x$,$\prtm x$
 respectively, and use eq.(13b) to obtain
 $$\Delta\Gamma^\mu_{\al\be}\prtp x^\al\prtm x^\be=\kp En^\mu\; .
 \eqno(16)$$ Thus,
 by eq.(8a), we find the equation of motion for $S$, (see appendix
 A for notation)
$$\nab x=-{\kp\over 2}\prtp x\times\prtm x\eqno(17)$$
 Repeating the above derivation in $\hat x$ coordinates, we find the
 corresponding equation for $\hat x(u)$:
$$\nab \hat x={\kp\over 2}\prtp \hat x\times\prtm \hat x\eqno(18)$$
 We shall refer to eqs.(17),(18) as the
  {\it vector} equations of motion;
 there is another, scalar equation, which we now derive.
  Eqs.(17),(18) describe the motion of the string under its own
 gravitational field. To render this picture consistent, we must
determine $P^{(t)}$ geometrically, and impose its equality to
 the expression (14) found from the gravitational field equations.
 On the top side of $S$, $\hat\Gamma=0$, and $\Gamma$ is therefore
 known from the transformation law of the connection. This readily
 gives:
 $$P_{\mu\nu}=
 {\prt\over{\prt n}}({{\prt\hat x^\lm}\over{\prt x^\mu}})
 {{\prt\hat x_\lm}\over{\prt x^\nu}}+
 {\prt\over{\prt n}}({{\prt\hat x^\lm}\over{\prt x^\nu}})
 {{\prt\hat x_\lm}\over{\prt x^\mu}} \eqno(19)$$
   From the definition of $\prt/\prt n$ (below eq.(9b)), we have
$$\prtpm({{\prt\hat x}\over{\prt n}})=\prtpm n^\mu{{\prt\hat x}
\over{\prt x^\mu}}+\prtpm x^\mu{\prt\over{\prt n}}({{\prt\hat x}\over
{\prt x^\mu}}) \eqno(20)$$
  From eqs.(19),(20) and the expressions\foot{With $k$ set to its
 on shell value, $-{\kp\over 2}$ (eq.(B.14)).}
 (B.5) for $\prtpm n$, we can find $P^{(t)}$; equating it
 to eq.(14), we get the conditions
 $$\prtpm\hat x^\mu\prtpm({{\prt\hat x_\mu}\over{\prt n}})=-n\cdot
 \prtpm^2x\; ,\eqno(21a)$$
 $$\sum_\pm\prtmp\hat x^\mu\prtpm({{\prt\hat x_\mu}
 \over{\prt n}})=-\kp E\; .\eqno(21b)$$
 The left-hand sides of eqs.(21) can be found by applying $\prtp$,
 $\prtm$ to eq.(9b), and using eq.(18). The result is that eq.(21b)
  is an identity, whereas eq.(21a) becomes the additional,
  {\it scalar}
 equation of motion (actually two scalar equations):
$$n\cdot\prtpm x=\hat n\cdot\prtpm\hat x \eqno(22)$$
 This equation, like the continuity condition (6), relates $x(u)$
 to $\hat x(u)$.
 \par The dynamics of the worldsheet, and therefore of spacetime,
 is determined
  by the orthonormality conditions (5), the continuity
 condition (6), the equations of motion (17),(18),(22) and the
 boundary conditions (1) and (2). For a {\it closed} string, the
 treatment is unchanged, except for the boundary conditions. As usual
 in string theory, we would then scale $u^a$ so $x(u)$,$\hat x(u)$
 are periodic in $\sigma$, with period $2\pi$. We still need to choose
 an arbitrary point $P$ on the string,
  so that the coordinate-transition region
 runs from $P$ to spatial infinity (see discussion in section 2). We
 may choose $\sigma=0$ at $P$. Then eqs.(1)-(3) still apply, if
 one understands $Q$ to mean the line $\sigma=2\pi$. $S$ has the
 topology of a cylinder, and we assume that the region of spacetime
 interior to it has the topology of the interior of a cylinder
\foot{Also, we again assume that $S$ does not self-intersect.}
 \foot{Ref. 8
  also considers the case where a section of
  the interior metric has the topology of a punctured disk.}. \par
  For a single, free point mass, equal-time sections of spacetime
 in a rest frame of the mass are
 cones, with the deficit angle (in our units of $8\pi G=1$) equal to
 the mass. In the case of our self-gravitating (open) string, the
 mass at each end-point, if nonzero, has the same
 geometrical interpretation\foot{ We impose the requirement that
  the end-masses be non-negative; the origin of this constraint
 is not geometrical, but rather physical --- to ensure energy
 positivity at the ends of the string.}.
  Alternatively, the endmasses may be
 {\it physically} defined, either as the strength of the boundary
 $\delta$-function terms that must be added to eq.(11) to ensure
 $T_{\mu\nu}$ conservation, or as Lagrangian parameters in an action
 formulation (see eq.(43) below). At first glance, it is
 not clear that the geometrical definition yields $\tau$-
 independent masses at $P$ and $Q$, but this is indeed the case, as
 proven in appendix C. The endmasses must satisfy some other
constraints, discussed in section 5.
\chapter{Integration using the Liouville Mode.}
 We now show how the dynamics of the self-gravitating string can be
 largely reduced to solving a single nonlinear wave equation on the
 worldsheet.
 \par Let us define the induced Liouville mode, $\phi$, as follows:
($E<0$ by eq.(7))
 $$ E(u)=-e^{\phi(u)} \eqno(23) $$
 where $E$ is the induced orthonormal scale factor (eq.(6)).
 As shown in appendix B (part II), the equations of motion imply
 the following differential equation for $\phi$ (eq. (B.22)):
 $$ \nab\phi+se^{-\phi}={\kp^2\over 4}e^\phi \eqno(24) $$
 where $s$ is a sign which, at least locally on the worldsheet, is
 undetermined. It is important to note that eq.(24) does not
 necessarily hold in the worldsheet coordinates defined below
 eq.(6). That is, (24) always holds locally for {\it some} orthonormal
 parametrization $u^a$, but in general, this will not be the same
 $u^a$ for which $\sigma=\; const$ at $P$ and $Q$\foot
{For the closed string, however, the functions $\be_+(u_+)$,
 $\be_-(u_-)$ which govern the requisite reparametrization (see
  eqs.(B.18)-(B.19)) are themselves periodic with period $2\pi$,
 so for closed string the new orthonormal coordinates respect the
 boundary conditions.}.

  In the flat case $\kp=0$, (24) reduces to the
 Liouville equation
  when $s=-1$. In the general case, eq.(24) is
 not soluble in closed form.
 It appears surprising, at first glance, that $\phi$ satisfies
 the Liouville equation in the flat case. But actually, this is
 a straightforward consequence of the equation of motion, which becomes
 $\prtp\prtm x=0$, and of the three-dimensionality of target space;
 Let us demonstrate this fact. Since $\prtpm x$ are null vectors and
 $\prtpm x$,$n$ span three dimensional space, it is easy to see
 that $$\prtpm x\times\prtpm^2 x=\be_\pm\prtpm x$$
 and $\be_\pm$, defined in (B.5b), depends only on $u^\pm$, since
 so does $\prtpm x$\foot{A less trivial fact, proven in appendix B,
is that $\prtmp\be_\pm=0$ even for $\kp\not=0$.}.
  Clearly $$\nab E=\prtp^2 x\cdot\prtm^2 x\; .$$
 Now, $\prtpm^2 x$ must be a linear combination of the two vectors
  $\prtpm x$ and
 $n$, so we easily find $$E\nab\ln(-E)=(\prtp^2 x\cdot n)
  (\prtm^2 x\cdot n)=\be_+\be_-,$$
 and by locally setting $\be_\pm$ to
  the signs $s_\pm$ via appropriate
 reparametrizations (that preserve orthonormality), we recover
 eq.(24) in this (flat) case.
 \par Once the Liouville mode $\phi$ is known, we can (in principle)
  solve for the
 entire configuration $\{x^\mu(u)\}$,$\{\hat x^\mu(u)\}$, as
 follows\foot{We describe the procedure for the bottom worldsheet;
  the determination of $\hat x(u)$ proceeds in the same way.
   In addition, in what follows we ignore boundary conditions
   and global aspects of the solution.}.
  Since $\be_\pm=s_\pm$, we have by virtue of eqs.(8) and (B.10),
$$\prtpm x\times\prtpm^2 x=\pm s_\pm\prtpm x \eqno(25a)$$
 In appendix B (part III), it is shown how to express $x^\mu(u)$ in
terms of two functions on $S$, namely
 $\gamma_\pm(u)$. (See eq.(B.25)).
We can `almost' determine these two functions: they satisfy the
equations
$$\prtpm\gamma_\mp=-{\kp\over 2}Es_\mp\prtmp\gamma_\mp, \eqno(25b)$$
which tell us the {\it directions} of the worldsheet vectors
$\prt_a\gamma_+$,$\prt_a\gamma_-$ at each point $u^a$ on $S$.
 This can be used to write a nonlinear first-order differential
 equation for the function $u^-=u^-(u^+)$ describing any
  $\gamma_+=\; const$
 line on the worldsheet, and similarly another such equation for
  $\gamma_-=\; const$ lines. (These differential equations are
 not in general soluble in closed form.). Once the
  $\gamma_\pm=\; c_\pm$ lines are known ($c_\pm$ constants),
   say in the form $u^-=f_\pm(u^+;c_\pm)$, we can solve for
 $c_\pm$ as functions of $u^a$, and this gives a {\it particular}
 solution of eqs.(25b); let us denote it by $\Gamma_\pm(u)$.
  The general solution of (25b) is then,
 $$\gamma_\pm=h_\pm(\Gamma_\pm(u)) \eqno(26)$$
 where $h_\pm$ are two arbitrary functions of a single argument; the
 eq.(B.28) then furnishes a functional equation for these two unknown
 functions \foot{Since eq.(B.28) holds for any $u^+$,$u^-$, it should
 in general be powerful enough to determine two functions of a
 {\it single} variable.}.
 Thus, we see that knowledge of $E(u)$ `almost' determines the
 configuration $x^\mu(u)$, in the sense that the remaining equations
 (two first-order differential equations and one functional equation)
 have, as unknowns, functions of a {\it single}
   variable. The equations
 which yield $\hat x^\mu(u)$ from $E$ are the same, except that $\kp$
 is replaced by $-\kp$.
\chapter{Global Structure and Asymptotic Form of Three-Metric.}
 In four-dimensional
 gravity, and for a physical system sufficiently localized in space
 to have an asymptotically Minkowskian spacetime at spatial infinity,
  the ADM procedure$\ll 10\rr$
  can be used to define and compute the total
  momentum
 and angular momentum of the system, in terms of the approach to
 asymptotia. A similar definition can be employed in 2+1 dimensional
 gravity, but with two differences:    \par
  I. Firstly, an `asymptotic' form
 of the metric holds {\it exactly} in the region exterior to any
  world-tube that contains the support of $T_{\mu\nu}$\foot{Such a
 worldtube clearly exists for our system.}. By `asymptotic form' we
 mean a standardized metric, depending on a finite number of degrees
 of freedom, and in a {\it single} coordinate patch\footnote{\dagger}
 {{\it Neither} of the three coordinate systems $\{z\}$,$\{x\}$ and
 $\{\hat x\}$ is an example of such a coordinate system.}
  which encompasses the entire
 exterior of the said worldtube. \par
 II. Such an asymptotic form does {\it not} approach a Minkowski
  metric at spatial infinity. \par
 The standardized asymptotic form depends on two scalar constants.
 This can be understood as follows\footnote{*}
 {For a more thorough treatment of the issue of asymptotic
 spacetime in three-dimensional gravity, including the implications
 for a canonical formulation (which is relevant to our sections 8
 and  9), see Brown and Henneaux$\ll 11\rr$ .
 These authors allow a cosmological
 constant.}:
  In the region exterior to the worldtube, consider a
   Minkowskian coordinate system, such as our $\{z\}$.
  Since the exterior is not simply connected,
 transition functions are needed from $M$ to itself when travelling
 around a noncontractable  loop; these functions are a Poincare
 transformation, which is just eq.(2). The Lorentz transformation
 $\bar L$ can either be boosted to become a pure rotation, in which
 case we will call it `rotationlike', or it cannot --- in the latter
 case it is `boostlike'. The boostlike case is the Gott regime, where
 arbitrarily large closed timelike curves occur
 in the exterior region; we will return
 to this case later. Here, let us restrict attention to the
 nonpathological, rotationlike case. By combining a shift with a
 boost, we can then choose new Minkowski coordinates in which
 $$\bar L=R(2\pi(1-a)),\; \bar b=(2\pi\be,0,0)\; ,\eqno(27)$$
 where $R(\al)$ denotes a spatial rotation by angle $\al$. This is the
 simplest form to which eq.(2) can be brought, and the following
 single-patch exterior metric encodes\footnote{\dagger}
 {Choosing the transition surface
  $T$ to run along $\varphi=\; const$, the coordinates
  $(t,r,\varphi)$, valid throughout the exterior, are related
 to the Minkowski polar coordinates $(t_m,r_m,\varphi_m)$ by
 $$r_m=r,\;\varphi_m=\varphi,\; t_m=t+\be\varphi\; .$$
 The exterior region with metric eq.(28) is causal, provided this
region is chosen to lie outside $r^2=\be^2/a^2$.}
  it:
 $$ds^2=-(dt+\be d\varphi)^2+dr^2+a^2r^2d\varphi^2\; .\eqno(28)$$
 This can be taken as the standardized asymptotic metric, for the
 case of rotationlike $\bar L$. \par
  Returning to a general Minkowski frame, the
  three-dimensional Lorentz transformation
 $\bar L$ is described by a three-vector $\bar w^\mu$, as follows:
  $$(\bar L)_{\mu\nu}=(1+{1\over 2}\bar w^2)\eta_{\mu\nu}-
 \sqrt{1+\bar w^2/4}\;\bar w^\lm\eps_{\mu\lm\nu}-{1\over 2}\bar w_\mu
 \bar w_\nu \eqno(29) $$
 $\bar L$ is rotationlike (boostlike) when $\bar w^2$ is negative
 (positive). The three-vectors $\bar w$ and $\bar b$,
  which characterize
 the exterior metric, are the general-relativistic generalizations
 of the flat-space concepts of {\it momentum} and {\it angular
momentum}, respectively; in the next section we will see that the
 two sets of parameters are in fact proportional to each other in
 the flat limit\footnote{\bullet}{This interpretation of
  the transformation eq.(2),
  was first elucidated in ref. 4.}.  \par
Finally, we briefly discuss other global aspects of the
 Nambu-Goto-Einstein (NGE) spacetimes. As pointed out in section 2,
 the global construction of these spacetimes from
$\{x(u),\;\hat x(u)\}$
(which was described in sections 2-3) breaks down if the string
intersects itself. Thus, some constructions
$\{x(u),\;\hat x(u)\}$ might not be physical; conversely, physical
spacetimes with self-intersecting strings will not be globally
 reproduced by our construction.\par Furthermore, even with no
 self-intersections, some constructions do not result in a spacetime
 with the assumed $R^3$ topology. As an illustration, consider two
 free, static masses with zero string tension$\ll 4\rr$
\footnote{*}
{This special case is dealt with in section 6.}. The top
and bottom copies of the surface $S+T$ in $\{z\}$ coordinates, may
be chosen as broken planes. Equal-time sections of two such
spacetimes are depicted in Figs.3a and 3b, which are special cases
 of Fig.2. In both
cases, the deficit angle at $P$ satisfies $$0<m^{(P)}<\pi\; .$$
But at $Q$, Fig.3a shows the case $0<m^{(Q)}<\pi$, whereas in Fig.3b,
 $\pi<m^{(Q)}<2\pi$. It is easily seen that the excessive deficit
 angle at $Q$ closes up the spatial section, so the spacetime of
Fig.3b does not have the topology $R^3$. Returning to the case
$m^{(P)}=m^{(Q)}$, the full constraint when $\kp=0$ is clearly
 $0\leq m\leq\pi$. For nonzero $\kp$, the upper bound may change,
although in any case $0\leq m<2\pi$. The upper bound is
 configuration dependent, and is determined by the requirement that
 the top copy of $S+T$ will not turn back on itself to intersect the
bottom copy (as occurs in Fig.3b).\par This test extends to a general
criterion, which determines whether some particular construction
$\{x(u),\;\hat x(u)\}$ corresponds to a nonintersecting NGE space
of topology $R^3$. The criterion is that {\it the two copies of
$S+T$ are only allowed to cut themselves or each other in ways that
do not close off any region of the section (Fig. 2)
 that is finite in
 $\{z\}$ coordinates.} \par This criterion imposes constraints, both
 on the functions $x(u)$,$\hat x(u)$ and on $\bar L$. When $\bar L$
is rotationlike, the latter constraint is (in the frame for which
 eq.(27) holds) $$a>0\; .$$ In the special case of Figs.3 (static
masses and $\kp=0$), this simplifies to the condition $m<\pi$.\par
Note that the above criterion {\it allows} sections to have the
 structure of a plane with multiple Riemann sheets. Any intersection
of a copy of $S$ with itself, that conforms to the criterion, does
{\it not} correspond to a self-intersection of the physical string.
\chapter{Perturbation Theory and Special Limits.}
 The developments in this section depend on the fact that, for
 small $\kp$, the equations of motion and boundary conditions
 may be solved in a
 perturbative series. In what follows, we shall restrict ourselves
 to the first order in $\kp$. All components in this section are
in $\{z\}$ coordinates continued around $P$ (see section 2).
 \par We solve the orthonormality condition and
 eq.(17) by expanding ($b$ is a constant three-vector),
$$x(u)=b+A(u^+)+\tilde A(u^-)-{\kp\over 2}A(u^+)\times\tilde A(u^-)
 +O(\kp^2) \eqno(30a)$$
  where the vector functions
   $A$,$\tilde A$ are arbitrary, except that
$$(A^\prm)^2=(\tilde A^\prm)^2=0. \eqno(30b)$$
  For $\kp=0$, this
 reduces to the standard flat-string separation into left- and
 right-moving modes. \par
 $\hat x$ can be expanded in the same form, with $A$,$\tilde A$
 replaced by different functions $C$,$\tilde C$;
 $b$ replaced by $\hat b$;
  and $\kp\rightarrow -\kp$. However, due to the
            continuity
 condition eq.(6), the scalar product $A^\prm(u^+)\cdot\tilde
 A^\prm(u^-)$ must be the same as its top side counterpart,
  to $O(\kp^0)$. Therefore we must have
 $$L\cdot C(u^+)=A(u^+)+const.+O(\kp)\; ,\eqno(31a)$$
 $$L\cdot\tilde C(u^-)=
 \tilde A(u^-)+const.+O(\kp)\eqno(31b)$$
 where $L$ is a constant Lorentz transformation.
 \par From eqs.(5),(6),(18) and (31), we find to order $\kp$:
 $$\eqalign{
 L\cdot\hat x(u)&=\hat b+rA(u^+)+{1\over r}\tilde A(u^-)-\kp
 B(u^+)-\kp\tilde B(u^-) \cr&+{\kp\over 2}A(u^+)\times\tilde A(u^-)
 +O(\kp^2)\cr}\eqno(31c)$$
 where $r=r(\kp)=1+O(\kp)$ is a constant to be determined, and
$$B^\prm=-A\times A^\prm,\;\tilde B^\prm=\tilde A\times\tilde A^\prm.
 \eqno(31d)$$
 The functions $B$,$\tilde B$ are determined up to constant additive
 three-vectors, which can be absorbed into $\hat b$. \par
 Unlike the Lorentz transformation $\bar L$ (eq. (2)), which is a
 physical observable, $L$ is not: it is easy to see that $x$,$\hat x$
 are unchanged under an order-$\kp$ change in $L$, if accompanied by
 suitable $O(1)$ shifts in $A$,$\tilde A$, $b$ and $\hat b$.
  \par
 In order to determine $r(\kp)$, it is necessary to impose the scalar
 equation of motion, eq.(22). Using eqs. (8a),(8c) as well, we find
 that $$r(\kp)=1+O(\kp^2)\eqno(31e)$$. \par
 As shown in appendix B (part (II)), the content of the scalar
 equation of motion is only a single
  numerical condition, once the vector
 equations of motion are used; so it is not surprising we are able
 to adjust the constant $r(\kp)$ so that eq.(22) is perturbatively
 satisfied.
  \medskip
 {\bf Boundary Conditions.} Specializing to the case of open string,
 we next impose on the perturbative solution (eqs.(30)-(31)) the
 boundary conditions--- eqs.(1),(3). We will only treat (1) in
  detail, since the handling of the $Q$ boundary condition
 proceeds in exactly the same way as that at $P$. \par
 Eq.(1) reads, after differentiating it w.r.t. $\tau$,
 $$(L-1)\cdot(A\upm+\tilde A\upm)-{\kp\over 2}(L+1)(A\times\tilde
 A)\upm=\kp(A\times A\upm-\tilde A\times\tilde A\upm)+O(\kp^2)\;
 ,\eqno(32a)$$
  where the arguments of $A$,$\tilde A$ are $\tau$, since $\sigma
 =0$ at $P$.
\medskip {\bf The Flat Limit.} This is the limit in which
 the end-masses $m$ tend to zero, while the ratio $m/\kp$ is held
 fixed. (For simplicity, we will assume throughout that the two
  endmasses
  are equal.). This includes the special case of the standard
 open string, for which the ratio vanishes; but we will emphasize
 here the case of massive end-points.
\par Since an endmass corresponds to a deficit angle (see section
 3), we expect the matrix $L-1$ to be of order $m$. Thus, the flat
 limit corresponds to taking $L\rightarrow 1$ while holding
 $(L-1)/\kp$ fixed. Letting
  $$L\cdot a\approx a-w\times a\; ,\eqno(32b)$$
  eq.(32a) becomes
$$(w-\kp(\tilde A-A))\times(A\upm+\tilde A\upm)=O(\kp^2)\eqno(33a)$$
 and therefore
$${{w^\mu}\over\kp}+\psi(A\upm+\tilde A\upm)^\mu\approx
\tilde A^\mu-A^\mu \eqno(33b)$$
 where $\psi$ is a scalar function of $\tau$. We decompose
 $${{w^\mu}\over\kp}\equiv\bar\rho\bar c^\mu,
 \;\;\bar c^2=-1\eqno(33c)$$
and set $\psi=-\bar\rho\bar\varphi$. In the flat limit, $\bar\rho$
and $\bar c$ are finite, and eq.(33b) becomes
$$\bar\varphi(A\upm+\tilde A\upm)=\bar c-{1\over{\bar\rho}}(\tilde A
-A)\; .\eqno(33d)$$
This implies that the square of the vector $\bar\varphi(A\upm+
\tilde A\upm)$ is $\tau$-independent, and thus $\bar\varphi=\bar\be
\varphi$, where $\bar\be$ is some constant and $\varphi(\tau)$
 is given by eq.(D.9b). Thus eq.(33d) is exactly (D.9a), with the
 identifications
  $$c^\mu=\bar c^\mu/\bar\be,\;\rho=\bar\rho\bar\be,
\eqno(33e)$$
  and we have in the flat limit
    $${w^\mu\over\kp}\rightarrow\rho c^\mu\; .\eqno(34a)$$
 We have thus recovered the flat boundary condition at $P$. In
 the corresponding procedure at $Q$, the arguments of $A$,$\tilde A$
 are $\tau+\pi$,$\tau-\pi$ respectively, and the matrix $L$ is replaced
 by $$L^{(Q)}\equiv L(\bar L)^{-1}\; ;\eqno(35a)$$
  In the flat limit,
 $$L^{(Q)}\cdot a\approx a+w^{(Q)}\times a\eqno(35b)$$
 for any vector $a$, and the counterpart of eq.(34a) is
 $${{w^{(Q)\mu}}\over\kp}\rightarrow\rho^{(Q)}d^\mu\; .\eqno(34b)$$
 For the flat string, $\rho=m^{(P)}/\kp$ and
  $\rho^{(Q)}=m^{(Q)}/\kp$, so
 these two numbers are equal when (as we assume) $m^{(P)}=m^{(Q)}=m$:
 $$\rho=\rho^{(Q)}=m/\kp\; .\eqno(34c)$$
  The constant vectors $c^\mu$,$d^\mu$
 are the three-velocities of $P$,$Q$ respectively, in the limit
 of vanishing flat-string tension ($\rho\rightarrow\infty$).
  We conclude that the boundary conditions at both $P$ and $Q$ reduce
 to their correct flat limits. The equations of motion themselves simply
 become $$x(u)=\hat x(u),\; \nab x=0,$$ which is the standard string
 equation of motion in orthonormal gauge.
\medskip {\bf Total Momentum and Angular Momentum.}
 As discussed in section 5, and as first found in ref. 4,
 the Lorentz transformation $\bar L$ and the shift $\bar b$  are
 closely related in the flat limit to the system's total momentum
 and angular momentum, respectively. We use the perturbative results
 to make this statement more precise. \par
   From eqs. (29),(32a),(34)-(35) we obtain
 to first order in $\kp$,
  $$\bar w^\mu\approx m(c^\mu+d^\mu)\eqno(36)$$
  On the other hand, the flat total momentum of
   the string is,(eq.(D.3))
 $$p_{total}^\mu=\kp\int_0^\pi d\sigma{{\prt x^\mu}\over{\prt\tau}}+
 m{{dx^\mu}\over{ds}}(P)+m{{dx^\mu}\over{ds}}(Q)\; .\eqno(37)$$
 Finally, in the flat limit eq.(30a) becomes
 $$x(u)=A(u^+)+\tilde A(u^-)\eqno(38)$$
  By using eqs.(D.5),(D.9) and the corresponding equations
  at $Q$, together with eqs.(36)-(38), we obtain
 to first order in $\kp$,
 $$\bar w^\mu\approx p_{total}^\mu \eqno(39a)$$
 in our geometrical units. Slightly more involved, yet
  straightforward, algebra yields for the shift vector:
 $$\bar b^\mu\approx J_{total}^\mu\; ,\eqno(39b)$$
 where the three-vector $J_{total}$ is dual to the total
  flat angular-momentum
 tensor of the system:
 $$(J_{total})_\mu=\eps_{\mu\al\be}\{
 \kp\int_0^\pi d\sigma x^\al
{{\prt x^\be}\over{\prt\tau}}+\sum_{P,Q}m(x^\al{{dx^\be}\over{ds}})
\} \; .\eqno(39c)$$
  Here $ds$ is an interval of proper time at either (massive)
  string-end.
 The above analysis can be repeated for the open string with lightlike
 endpoints (the case $m$=0), or for a closed string, and one obtains the
 same relations (39a)-(39b). In both cases, $J_{total}$ and
  $p_{total}$ will
 not contain any boundary terms.

\medskip {\bf Zero Tension Limit.} This we define to be the limit
 where $\kp\rightarrow 0$ with $L$,$L^{(Q)}$ fixed; actually, we
  shall consider a small but nonvanishing $\kp$, so that the
 above perturbative expansion may be used. Thus, the string tension
 is small in geometrized units\foot{Since $\kp$ has inverse-length
 dimensions in these units, what is meant is that $\kp D<<1$,
  for some length or time scale $D$
 typical of the system.}, but the endmasses give
 rise to finite deficit angles in their respective rest frames. The
 boundary condition at $P$ is again given by eq.(32a), where now
 $(L-1)$ is of order $1$. As usual, a corresponding condition holds
 at $Q$. \par
These boundary conditions can be solved in power series in $\kp$;
 the procedure is quite similar to the large-$\rho$ expansion of the
 flat string, described in appendix D, and will not be presented here.
  When $\kp=0$, the solution of eq.(32a) is $A\upm(\tau)+\tilde A\upm
(\tau)\vert\vert w$; this is equivalent to
 the $\rho\rightarrow\infty$ limit of the flat string,
 and implies that the worldline $P$ is rectilinear\foot{See comment
 below eq.(D.9b).} --- it is a free
 mass (as is $Q$). Since the rest-frame deficit angles at $P$ and $Q$
 are $m$, it is easy to see that
 $$w_\mu=2\sin{m\over 2}c_\mu,\;\; w_\mu^{(Q)}=2\sin{m\over 2}
 d_\mu\; ,\eqno(40)$$
where $L$ is related to $w^\mu$ via the same relation (eq.(29))
 as holds between $\bar L$ and $\bar w^\mu$, and
  $(L^{(Q)})^{-1}$ is again related in this way\foot{Eqs. (32),
(35b) are the linearized versions of these relations.}
   to the vector $w^{(Q)}$. The vectors $c$,$d$ are the constant
   three-velocities of $P$ and $Q$, respectively. $m$ must satisfy
 $0\leq m\leq\pi$ to ensure the $R^3$ topology of spacetime (see
  section 5).  \par
 Rearranging eq.(35a), we have $$\bar L=(L^{(Q)})^{-1}L\eqno(41)$$
  From eqs.(40)-(41), we can determine the relative velocity that
 must be imparted to the two masses in order to enter the Gott regime:
 specializing the $\{z\}$ coordinate system to a center of mass frame,
  $c^0=d^0$,$c^i=-d^i$ and we
  obtain:
 $$\bar w^2=4(w^0)\ll{{(w^0)^2}\over 4}-1\rr \eqno(42)$$
 The vector $\bar w$ becomes spacelike, and $\bar L$ `boostlike',
 when $w^0>2$; In this regime, closed timelike curves exist
 sufficiently far away from the two masses\foot{ Assuming that the
 two masses have a nonvanishing impact parameter, `sufficiently far'
 means relative to the impact parameter. The CTC's must also have a
 sufficiently large extent, in center-of-mass time, around the
 fiducial time --- the instant in which the masses are
 at minimal distance from each other.}, and there are CTC's of
  {\it arbitrarily} large extent in space and time.\par
 On the other hand, when $\bar w$ is timelike, {\it no} CTC's or
 CLC's exist in the zero-tension
 limit, and spacetime is completely causal$\ll 5\rr$. \par
 When a small, nonzero string tension is turned on,
the $\kp$-perturbative corrections do not change these causal
 properties of the Gott spacetime.
 We thus conclude that slightly
 accelerating the two particles, at least via the mechanism of
 tying them to opposite ends of a Nambu-Goto string, {\it cannot}
 produce CTC's in a previously-causal spacetime, since a spacetime
 with arbitrarily large CTC's (in the above sense) cannot be initially
 causal.

 \chapter{Causal Structure of Spacetime.}
 We have assumed that\foot{See discussion in section 2.} some
neighborhood of the worldsheet $S$ has spacelike
equal-time sections; such a
 neighborhood has a normal causal structure --- that is, no closed
 timelike/lightlike curves exist within it. \par
 The entire spacetime manifold, on the other hand, may have closed
 timelike curves (CTC's) or closed lightlike curves (CLC's),
 although the energy-momentum tensor (eq.(11), plus the boundary
 contributions) satisfies all standard energy-positivity
 criteria\foot{
 These are the {\it weak, strong} and {\it dominant} energy
  conditions$\ll 9\rr$.}
 and there are no horizons\foot{The two-wormhole `time machine'
 of Morris {\it et al}$\ll 12\rr$ is an example
  of a $3+1$-dimensional spacetime
 with CTC's, in which the weak energy condition is violated. The
 Kerr-Newman black hole with $a^2+e^2\leq m^2$ ($a\not=0$) has
CTC's, but they
are hidden behind an event horizon$\ll 13\rr$ .}.
  An example is when $\kp<<1$ and $\bar L$ is boostlike (the Gott
  regime--- see section 6). In that case arbitrarily large
 \foot{in both space and time; see last footnote of section 6.}
 CTC's exist, but arbitrarily {\it small} CTC's do not exist.
 Another example of such a spacetime (also in $2+1$ dimensions) is
 Godel's solution$\ll 9\rr$ for a homogenous dust
 universe. However, one might take the attitude
 that such spacetimes should be legislated away. A more interesting
 structure is one in which the energy conditions are satisfied,
  there are `naked' CTC's or CLC's\foot{That is, outside any
   horizon.}, {\it and} these curves do not extend before a finite,
 globally-defined time. In such a spacetime, the CTC's are `produced'
 in a universe which was, until that epoch, causal
 \foot{The significance of the ability to {\it produce} CTC's at
 finite time, was emphasized to me by Lenny Susskind.}.
  Familiar examples of such spacetime are the four-dimensional
  empty Taub-NUT space,
 and its two-dimensional analog --- the Misner space$\ll 9\rr$
 \foot{These two manifolds, however, are geodesically incomplete
 near the CTC-production epoch (and at the time when CTC's end).
 Geodesic incompleteness (timelike or null) is a kind of
  singularity, in that it dooms some observers to finite affine
 lifetimes.}. \par
  If some physical process is found to generate CTC's
 which are accessible to (macroscopic or microscopic) observers,
 it will become important to ascertain the physical ramifications of
 these curves.
 Back-reaction effects might conspire to prevent CTC's from forming;
 alternatively, it might become necessary to study self-consistent
 propagation of fields in spaces with CTC's$\ll 14\rr$ .
 \par
 Returning to the self-gravitating string, an interesting question
 is whether there exists
   a regime for which CTC production can occur, in a previously
 causal spacetime. As we learned in the previous section, such
 solutions can only exist, if at all, for large $\kp$. Also, such
 a solution must have a rotationlike $\bar L$, since otherwise some
 CTC's are {\it guaranteed} to extend into past
  infinity, just as for the
 Gott solution. We do not yet know whether there are such
 CTC-producing solutions
  of the self-gravitating string\footnote
 {\dagger}{ In the absence of general theorems to the contrary, it
 remains possible that small accelerations applied to
 two four-dimensional cosmic strings might
 enable the artificial generation of CTC's, as Gott suggests. But
 as a result of our perturbative analysis in section 6, we know
 that (for small accelerations)
 either the acceleration mechanism is different from the
 one we have considered, or the four-dimensionality of spacetime
 would enter in a crucial way in such a procedure --- for
 instance, through the finite length of the cosmic string, or horizon
  formation. In a four-dimensional
   interpretation of our spacetime, the
 endmasses become cosmic strings, whereas {\it our} string should be
 re-interpreted as a membrane connecting them.}.
  \par
 Finally, we note that unusual causal structures are of interest not
only classically or semiclassically. Most of the work that has been
 done to date on quantum gravity uses, in the sum over spacetime
 histories, metrics of euclidean signature. An appeal is then made
 to some version of Wick rotation, in order to make statements
 about Minkowskian spacetime. It is thus important to attempt the
 direct study of path integrals over Minkowskian metrics, and such
an integral may need to include a sum over causal structures, as well
 as over topologies.
\chapter{An Action Formulation.}
 In this section we shall start with the Nambu-Goto-Einstein action,
 eliminate the nonpropagating metric, and end up with a quadratic
\footnote{*}
{The action is actually cubic if one views the Lagrange
 multipliers as fields.}
 action, the dynamical degrees of freedom all living on the
 worldsheet. These degrees of freedom are $x,\hat x,n$ and $\hat n$,
 and Lagrange multipliers are introduced to enforce geometric
  constraints, which are also quadratic in the worldsheet fields.
 For simplicity, we shall restrict ourselves in this section to
 spacetimes for which $\bar L$ is
  rotationlike (see section 7). \par
 The Nambu-Goto-Einstein action, in an orthonormal worldsheet
  parametrization, is the sum of the three-dimensional
 Einstein-Hilbert action, the bulk worldsheet action, and
 terms for the endmasses\footnote{\dagger}{The Lagrangian
  mass parameter $m$
 in this action, is the same geometrical parameter given
  on shell by eq.(C.1).}:
$$\eqalign{
S_{NGE}(x,\hat x,g)&=-{1\over 2}\int d^3xR\sqrt{-g}+\kp\int du^+du^-
g_{\al\be}\prtp x^\al\prtm x^\be \cr &
+\sum_{P,Q}m\int d\tau
\sqrt{-g_{\al\be}
{{dx^\al}\over{d\tau}}{{dx^\be}\over{d\tau}}}\cr}\eqno(43)
$$
The extremization of $S_{NGE}$ does not quite yield the gravitational
  equation of
 motion: the discrepancy stems from the surface terms in the variation
 of the first (Einstein-Hilbert) term, which arise when integrating
 this variation by parts. Such surface terms depend only on
 the asymptotic form of the metric (see section 5). In four
 dimensions, and for asymptotically Minkowskian spacetime,
 the asymptotic form\footnote{**}
 {Or rather, its leading deviation from the
 Minkowski metric.} is characterized by the total momentum of
 the system, and it is well known$\ll 10\rr$ how
 to modify the action and the Hamiltonian formalism to accomodate
 these degrees of freedom. Similarly, in three spacetime dimensions,
 the surface terms depend only on $\bar w$ and $\bar b$.
 \par In what
 follows we shall circumvent this issue by constraining
 the variation $\delta g_{\al\be}$ to be such as to preserve
 $\bar w$; as a consequence, the scalar equation (22) will
 not result from extremizing our action, and must be put in by hand
\footnote{\bullet}
{Recall (appendix B) that the extra content in this equation
 is just one numerical condition.} \foot{For a general
  analysis of the canonical formulation
 of three-dimensional gravity which treats in detail
 the asymptotic degrees of freedom, see ref. 11.}.
 Before leaving behind the issue of the surface terms, however,
 we make two
 observations. One is that although we fix a
  {\it vector} $\bar w^\mu$,
 only one scalar equation will be lost. The reason is that,
 by a suitable coordinate transformation,
 the asymptotic form can be chosen to be eq.(28), for which
 $\bar w$ has a single component $\bar w^0$ (see eq.(27)).
 In other words, two of the missing equations of motion are
 identities, thanks to three-dimensional general coordinate
  invariance. The other observation is that we need only fix
 $\bar w$ (not $\bar b$) in varying the metric, because
 the surface terms do not depend on $\bar b$. In fact, the
 surface terms are$\ll 9\rr$\footnote{\dagger}{
 Here $\Sigma$ is some cylindrical surface enclosing the worldsheet,
 and $d\sigma$ an outward-pointing normal to $\Sigma$
  with a magnitude equal to a surface-element area.}
$$-{1\over 2}\int g^{\al\be}\delta R_{\al\be}\sqrt{-g}d^3x=
{1\over 2}\int_\Sigma(g^{\al\be}\delta\Gamma^\gm_{\al\be}-g^{\gm\al}
\delta\Gamma^\lm_{\al\lm})d\sigma_\gm\; ,\eqno(44)$$
  and using the
 asymptotic metric (28), we find that (44) equals
 $$-2\pi\delta a\int dt$$
 which does not depent on $\be$. \par
 We are now ready to eliminate the metric from the formalism. In the
 coordinate system $\{x\}$, $g_{\mu\nu}$ is Minkowskian on the
 worldsheet, so it drops out of the last two terms in eq.(43).
 The first term can be computed as follows. First, the procedure
  of section 3
  (eqs.(19)-(22)) is repeated {\it off-shell}, i.e. assuming only
 orthonormality and metric continuity (that is, not using any
equations of motion). This yields $P^{(t)}$, which together with
 eq.(13b) finally gives
$$-{1\over 2}\int d^3x R\sqrt{-g}=
2\int du^+du^-(n\cdot\nab x-\hat n\cdot\nab\hat x)\; .\eqno(45a)$$
 To summarize, $g_{\mu\nu}$ is eliminated, purely geometrically, in
 terms of the top and bottom worldsheets, and when the action
 $S_{NGE}$ is expressed in terms of $x$ and $\hat x$, the result
 is: $$\eqalign{
 S_1(x,\hat x)&=2\int du^+du^-(n\cdot\nab x-\hat n\cdot\nab\hat x)
+{\kp\over 2}\int du^+du^-(\prtp x\cdot\prtm x \cr &+
\prtp\hat x
\cdot\prtm\hat x)+
\sum_{P,Q}m\int d\tau\sqrt{-({{dx}\over{d\tau}})^2}\cr}
\eqno(45b)$$
 where the second (bulk string) term has been rendered top-bottom
 symmetric by use of eq.(6), and $n,\hat n$ are given by eqs.(8).\par
In (45b), the configuration $x(u),\hat x(u)$ is free, except for
 the orthonormality and metric-continuity constraints, which we
 rewrite,
$$(\prtpm x)^2=(\prtpm\hat x)^2=0\eqno(45c)$$
$$\prtp x\cdot\prtm x=\prtp\hat x\cdot\prtm\hat x\eqno(45d)$$
 and the boundary conditions, eqs.(1)-(3), which we rewrite thus:
 $$x(P)=\hat x(P)\; ,\eqno(46a)$$
 $${d\over{d\tau}}x(Q)=\bar L\cdot{d\over{d\tau}}
 \hat x(Q)\; .\eqno(46b)$$
 We have replaced eq.(3) by its $\tau$-derivative, since as explained
 above, we are only holding $\bar w$ fixed, not $\bar b$, so the weaker
 constraint (46b) is appropriate. The bulk constraints\foot{We use
 the term {\it bulk} to refer to the interior of the worldsheet, i.e.
 away from the boundaries $P$ and $Q$.}, eqs.(45c)-(45d), and
 the end constraints eqs.(46), will be accounted for by adding to
 the action $S_1$ the appropriate Lagrange-multiplier terms. For
 the case of the closed string, the last term in $S_1$ is absent,
 and the boundary conditions eqs.(46) are modified as described
 in section 3.  \par
 The action (45b) is nonpolynomial in $x$ and $\hat x$, since
 $n,\hat n$ are; this makes it awkward to work with, and
 renders the task of quantization rather hopeless. Fortunately, this
 problem can be avoided if we elevate $n(u),\hat n(u)$ to the status
 of additional dynamical fields on the worldsheet; in this picture,
 however, we must impose the further bulk constraints,
 $$n^2=\hat n^2=1\; ,\eqno(47a)$$
$$n\cdot\prtpm x=\hat n\cdot\prtpm\hat x=0\; .\eqno(47b)$$
  all the constraints (eqs. (45c)-(47b)) are quadratic, and so is
 the action $S_1$ in the new picture --- except for the endmass
 terms\foot{That is not a problem, however, since we would only be
 interested in quantizing the self-gravitating string for
 a closed string, or for
  an open string with massless ends; in either case there are
 no endmass terms.}. \par
 The quadratic action, obtained by adding to $S_1$ all the
 constraints weighted by their Lagrange multipliers, is:
$$\eqalign{S_2 &=
2\int du^+du^-(n\cdot\nab x-\hat n\cdot\nab\hat x)+\int du^+du^-
\ll({\kp\over 2}+\lm_1)\prtp x\cdot\prtm x \cr &+
 ({\kp\over 2}-\lm_1)\prtp
\hat x\cdot\prtm\hat x\rr+\int du^+du^-\ll\sum_{\pm}\lm_2
^\pm(\prtpm x)^2+\sum_{\pm}\hat\lm_2^\pm(\prtpm\hat x)^2\cr &+
 \lm_3(n^2-1)
+\hat\lm_3(\hat n^2-1)+
\sum_{\pm}\lm_4^\pm n\cdot\prtpm x
+\sum_{\pm}\hat\lm_4^\pm\hat n\cdot\prtpm\hat x\rr
\cr &+\int d\tau\ll\mu_1(\tau)\cdot(\hat x(P)-x(P))+\mu_2(\tau)
\cdot({{dx(Q)}\over{d\tau}}-\bar L\cdot{{d\hat x(Q)}\over{d\tau}})\rr
\cr &\;\; +\sum_{P,Q}
 m\int d\tau\sqrt{-(dx/d\tau)^2}\cr}\eqno(48)$$
 The multipliers $\lm_1,\lm_2^\pm,\hat\lm_2^\pm,\lm_3,\hat\lm_3,
\lm_4^\pm,\hat\lm_4^\pm$ are scalars, whereas $\mu_1,\mu_2$ are
 three-vectors. \par
 In the remainder of this section, we shall ignore the
 boundary terms in $S_2$, and analyze only the bulk equations of
 motion resulting from varying the action (48). Before embarking
 upon this analysis, we briefly summarize the results.
  We shall find that
 these equations of motion are consistent with the correct
 vector equations (17)-(18). They must, in fact, be {\it equivalent}
 to them, since the extremization of $S_2$ is equivalent to that of
 the original $S_{NGE}$; from our direct analysis of (48), however,
 we were only able to show that this equivalence is plausible.
 We also find that
 the numerical parameter $r_1$ can be chosen arbitrarily, at least
 if one only varies the fields $x,n,\hat x,\hat n$ in the bulk of
 the worldsheet\foot{We have not yet checked how the
  worldsheet surface terms affect this statement, but it seems they
 will not determine $r_1$, since eq.(22) was derived from {\it local}
 analysis of the equations of motion.}.
  Thus, the scalar equation $r_1=1$ is not reproduced by variation
 in the bulk. As discussed in the beginning of this section, we
 attribute this missing information to the fact that $\bar w$ must be
 held fixed in the variation; the condition $r_1=1$ must be imposed
 by hand. \par
 We now proceed with the bulk extremization\foot{
  Integrations by part can be freely performed on the worldsheet,
  since we are ignoring boundary
 terms.} of $S_2$. \par
 Variation w.r.t. $n(u)$ gives,
$$2\nab x+2\lm_3 n+\sum_\pm\lm_4^\pm\prtpm x=0$$
so from orthonormality, $$\lm_4^\pm=0\; .\eqno(49a)$$
 Thus $\nab x=-\lm_3n$;
 whence, in the notation of appendix B, $$\lm_3=-kE\eqno(49b)$$
 Similarly, $\delta/\delta\hat n$ yields
 $$\hat\lm_4^\pm=0,\;\hat\lm_3=\hat k E\; .\eqno(49c)$$
 Extremizing $S_2$ w.r.t. $x$, we find
$$\nab n-({\kp\over 2}+\lm_1)\nab x-\sum_\pm\prtpm(\lm_2^\pm\prtpm x)-
{1\over 2}\sum_\pm(\prtpm\lm_1)\prtmp x=0\eqno(50)$$
  From this vector equation, three scalar equations are obtained, by
 dotting it with $\prtpm x$ and with $n$. Using the off-shell result
 (B.5), the $\prtpm x$ components of eq.(50) become:
 $$2\prtmp(\be_\pm-E\lm_2^\mp)=E\prtpm\lm_1\; ,\eqno(51)$$
 whereas the $n$ component is
$$\prtp n\cdot\prtm n=-({\kp\over 2}+\lm_1)kE+\sum_\pm\lm_2^\pm
(\prtpm n)\cdot(\prtpm x)\; .\eqno(52)$$
 Using eq.(B.5a) again, this becomes
$$-2kE^2f=\be_-g_++\be_+g_-\; ,\eqno(53a)$$
where we define $$f\equiv\lm_1+k+{\kp\over 2}\; ,\eqno(53b)$$
$$g_\pm\equiv\be_\pm-2E\lm_2^\mp\; .\eqno(53c)$$
 Next, use eqs.(B.8a) and (53b)-(53c) to rewrite (51) as follows:
$$\prtmp g_\pm=E\prtpm f\eqno(54)$$
 Extremizing $S_2$ w.r.t. $\hat x$,
  and following the same algebraic
 manipulations, we obtain
$$2\hat kE^2\hat f=\hat\be_-\hat g_++\hat\be_+\hat g_-\eqno(55a)$$
$$\prtmp\hat g_\pm=E\prtpm\hat f\eqno(55b)$$
where $\hat\be_\pm$ are defined in eq.(B.6b), and
$$\hat f\equiv -\lm_1-\hat k+{\kp\over 2}\; ,\eqno(55c)$$
$$\hat g_\pm\equiv\hat\be_\pm-2E\hat\lm_2^\mp\; .\eqno(55d)$$
 These are all the bulk Euler-Lagrange equations resulting from $S_2$,
 apart from the constraints themselves. \par
 We can immediately solve for $\lm_1,k,\hat k$ by inspection --- by
setting $$f=\hat f=k+\hat k=0\; .\eqno(56a)$$
 Then, by (53b) and (55c),
$$\lm_1=0,\; k=-\hat k=-{\kp\over 2}\; .\eqno(56b)$$
  Eqs.(54) and (55b) then imply,
 $$\prtmp g_\pm=\prtmp\hat g_\pm=0\; ,\eqno(57a)$$
 which together with (53a),(55a) gives:
$$\be_-/\be_+=-g_-(u^-)/g_+(u^+)\; ,\eqno(57b)$$
$$\hat\be_-/\hat\be_+=-\hat g_-(u^-)/\hat g_+(u^+)\; .\eqno(57c)$$
But by eqs.(B.8),(56):
$$\be_+=\be_+(u^+),\;\be_-=\be_-(u^-)\eqno(57d)$$
$$\hat\be_+=\hat\be_+(u^+),\;\hat\be_-=\hat\be_-(u^-)\eqno(57e)$$
 Therefore eqs.(57b)-(57c) determine $g_\pm,\hat g_\pm$ up to four
 numerical constants\foot{Once $g_\pm$ and $\hat g_\pm$ are known,
 the multipliers $\lm_2^\pm,\hat\lm_2^\pm$ are known from eqs.(53c)
 and (55d).}.  \par
 From this point on, the analysis of appendix B, part (II) applies,
 since we are on shell--- except that the constant $r_1$ is not
 determined, as mentioned above, and must be set to $1$ by hand.
 We know that eq.(56a) must follow from the equations of motion
 of the action $S_2$, but have not proven it directly,
  although it appears to us plausible that eqs.(49)-(55) indeed
 imply (56a).
\chapter{Quantization}
The action formulation of the last section is a promising point of
departure for quantizing the self-gravitating string. To do so, a
 Hamiltonian should be derived from the worldsheet action $S_2$.
 Then, Unless a particular orthonormal parametrization (such as
 the light-cone gauge in standard string theory) can be found where
 all the constraints are easily soluble, these constraints must be
 imposed as conditions on physical states --- as is done in
 the covariant formulation of standard strings\foot{Incidentally,
  the light-cone orthonormal
  gauge {\it cannot} be chosen here, since in
 three dimensions this gauge exists only when $\nab x=0$.}.
 \par Another possible starting point for quantizing the
 theory, is the classical Liouville-like equation (24).
  It is, of course, likely that other conformal
 matter must be added on the worldsheet to render the quantization
 consistent--- this is certainly the case for $\kp=0$. \par
 What could such a string theory mean?
 Even though it is a first quantized string, it automatically
includes the gravitational interactions amongst different
 portions of the string. These interactions do not include
 graviton exchange, since there are no gravitons in three-dimensional
 Einstein gravity\foot{Nor do they include any of the quantum physics
  of pure $3d$ gravity}.
 Therefore, scattering amplitudes constructed in this theory
 would still be tree diagrams, from the point of view of
unitarity. Yet, they would include effects to all orders in
 $\kp$ --- which means, in non-geometrized units, all orders in
 $G$. Thus, we expect the {\it first quantized}
  self-gravitating string
 to include some string field-theoretic effects --- perturbative, as
 well as nonperturbative\foot{However, if the quantization program
 can be carried out, it will certainly be simpler to extract $\kp$-
 perturbative corrections to flat (standard) string theory, than
 to solve the $\kp$-nonperturbative theory.}.
 This model could therefore be a laboratory for investigating
 string-nonperturbative physics in the continuum.
\chapter{Conclusions}

 We have investigated
 a family of classical spacetimes in $2+1$ dimensions, produced by
  a Nambu-Goto self-gravitating string. Due to the special properties
  of three-dimensional gravity, the metric is completely described
  as a Minkowski space with {\it two} identified worldsheets.
 The geometry of our spacetime is expressed as a flat region of
  Minkowski space, with the two
  worldsheets identified. The equations of motion of the worldsheet
 were found, and reduced to a Liouville-like equation for the
 induced worldsheet Liouville mode.
  The flat limit and the zero-tension limit were worked out
 for the case of open string with massive endpoints.
 We have shown how to expand solutions in the geometrized string tension.
 For small string tension, spacetime was found to have the causal
structure found by Gott.
  The two-worldsheet formalism was recast, using auxiliary fields,
   as an action principle on the worldsheet. The new action is
  quadratic (with quadratic constraints)
   and perhaps amenable to consistent quantization.
   We suggest that the {\it first}
   quantized self-gravitating string could be
 a step towards understanding nonperturbative string field theory in
 a continuum setting.
\par
    Further work is in progress, mainly to find the possible causal
  and global structures of our classical spacetimes\foot{Including
 the physics of self-intersecting strings.},
    and to ascertain whether
 the quantization scheme proposed here can be implemented.
\par
 We have concentrated mostly on the case of open string with
 massive endpoints. In a quantized theory, one should
 work either with standard open strings (i.e., lightlike ends)
 or with closed strings. The treatments differ only in the
 boundary conditions; we have indicated the required changes.
\medskip{\bf Appendix A: \ Conventions and Units.}\smallskip
 We use greek letters for three-dimensional world
  indices, which
 run from $0$ to $2$, and the vector component notation
 $a^\mu=(a^0,a^1,a^2)$; lower-case latin indices
  $i,j,...$ denote
 world (target-space)
 spatial indices, whereas $a,b,...$ are sometimes used to
  denote worldsheet indices.
 The Minkowski metric is $\eta_{\mu\nu}=(-1,1,1)$, and the totally
 antisymmetric
 $\epsilon$-symbol is defined by $\eps_{012}=1$. \par
 The Minkowski scalar product of two three-vectors is denoted
 $a\cdot b=a^\mu b_\mu$. We also define a vector product $a\times b$,
 thus: $$(a\times b)_\mu\equiv\eps_{\mu\al\be}a^\al b^\be\; .$$
 This vector product obeys the relation
 $$(a\times b)\times c=a(b\cdot c)-b(a\cdot c)\; ,\eqno(A.1)$$
  which has the
  opposite sign from the corresponding Euclidean relation. \par
We denote by $\nab$ the worldsheet d'Alembertian: $$\nab=\prtp\prtm$$
\par
 In general relativity, we employ geometrized units $8\pi G=1$, where
 $G$ is Newton's constant in $2+1$ spacetime dimensions, and conform
 to the curvature-sign conventions of Weinberg$\ll 15\rr$.
\medskip{\bf Appendix B:
 \ Differential Geometry and Equations of Motion.}\smallskip
 This appendix has three parts. In part (I) we derive some useful
 differential-geometric results {\it off-shell} --- that is, we only
 assume orthonormality and metric-continuity, but do {\it not} yet
 assume the equations of motion. This part is thus particularly
 useful for the action formulation.  \par
 In part (II), we impose the vector equations of motion, show that
 the {\it scalar} equation of motion, eq.(22),
  is then only a numerical
 constraint, and impose it as well. We then show that the induced
 Liouville mode satisfies the differential equation, eq.(24). \par
 Finally, part (III) is concerned with finding self-gravitating
 string solutions once the Liouville mode $\phi$ is known (see
 section 4 in text). Throughout this appendix, we ignore the
 boundary conditions. \par
 {\bf (I) \ Off Shell Results:} Consider a spacetime, constructed as
  described in section 2 (open or closed string). Disregarding the
 boundary conditions, we are left only with the orthonormality
 condition (eq.(5)) and the metric-continuity condition (eq.(6)).
 By applying $\prtmp$ to eqs.(5) and using eqs.(8a),(8c) we find,
 $$\nab x=kEn,\;\; \nab\hat x=\hat k E\hat n \eqno(B.1)$$
 where $k$,$\hat k$ are unknown scalar functions on the wordsheet.
 We have (eq.(8b)),
 $$n\cdot\prtpm x=0\; .\eqno(B.2)$$
 Applying $\prtpm$,
 $$\prtpm^2 x\cdot n=-\prtpm n\cdot\prtpm x, \eqno(B.3)$$
 whereas applying $\prtmp$ to (B.2) yields
 $$ \prtmp n\cdot\prtpm x=-kE\; .\eqno(B.4)$$
 Eqs.(B.3),(B.4) and $n\cdot\prtpm n=0$ (which holds since $n^2=1$),
 give all three components of $\prtpm n$ in the basis $\{\prtp x,
\prtm x,n\}$, and allow us to expand
 $$\prtpm n=-k\prtpm x+{1\over E}\be_\pm\prtmp x\; ,\eqno(B.5a)$$
  where $$\be_\pm\equiv -(\prtpm^2x)\cdot n\; .\eqno(B.5b)$$
 The above derivations follow through for the variables of
 the top worldsheet as well, so
 $$\prtpm\hat
  n=-\hat k\prtpm\hat x+{1\over E}\hat
  \be_\pm\prtmp\hat x\; ,\eqno(B.6a)$$
  where $$\hat\be_\pm\equiv -(\prtpm^2\hat x)
  \cdot \hat n\eqno(B.6b)$$
 Since the formulae for $\{\hat x,\hat n\}$ are in exact correspondence
 with those for $\{x, n\}$, we shall mostly work with the latter.
\par
The integrability condition of eqs.(B.5a) for $\prtpm n$ is:
$$(\prtm k)\prtp x-(\prtp k)\prtm x=
\prtm\ll{{\be_+}\over E}\prtm x\rr-\prtp\ll{{\be_-}
\over E}\prtp x\rr\; .\eqno(B.7)$$
Dotting this with $n$ produces an identity, by the second of
 eqs.(B.5); but dotting (B.7) with $\prtpm x$, and use of
 orthonormality, yields
 $$\prtpm\be_\mp=-E\prtmp k \eqno(B.8a)$$
 and similarly $$\prtpm\hat\be_\mp=-E\prtmp\hat k \eqno(B.8b)$$
 Next, we apply $\prtpm$ to eq.(6) and use eq.(B.1):
$$\prtpm^2x\cdot\prtmp x=\prtpm^2\hat x\cdot\prtmp\hat x
=\prtpm E \eqno(B.9)$$
 This, together with $\prtpm^2x\cdot\prtpm x=\prtpm^2\hat x\cdot
 \prtpm\hat x=0$ and eqs.(B.5b),(B.6b), gives the components
 of the four three-vectors $\prtpm^2x,\prtpm^2\hat x$ in the bases
 $\{\prt x,n\}$ and $\{\prt\hat x,\hat n\}$, respectively:
$$\prtpm^2x={1\over E}(\prtpm E)\prtpm x-\be_\pm n\eqno(B.10a)$$
$$\prtpm^2\hat x={1\over E}(\prtpm E)
\prtpm\hat x-\hat\be_\pm \hat n\eqno(B.10b)$$
 Now, apply $\prtmp$ to eqs.(B.10), and utilize (B.1): (recall $E<0$)
 $$\prtpm(kEn)=\nab\{\ln(-E)\}\prtpm x+(\prtpm\ln(-E))kEn-
 (\prtmp\be_\pm)n-\be_\pm\prt_\mp n \eqno(B.11)$$
 and a similar relation for $\hat x,\hat n,\hat\be_\pm$. Dot
 (B.11) with $\prtmp x$, and use eq.(B.5a) for $\prtmp n\cdot
 \prtmp x$ and $\prtpm n\cdot\prtmp x$; the result is
$$E\nab\ln(-E)=\be_+\be_--k^2E^2\; , \eqno(B.12a)$$
and similarly
$$E\nab\ln(-E)=\hat\be_+\hat\be_--\hat k^2E^2\; . \eqno(B.12b)$$
 Comparing the two eqs.(B.12), we find the relation
 $$\be_+\be_- -k^2E^2=\hat\be_+\hat\be_- -\hat k^2E^2
 \; .\eqno(B.13)$$
{\bf (II) \ The Equations of Motion:}
 On shell, we have at our disposal also the vector equations of
motion (17)-(18), and the scalar equation, eq.(22). We begin by using
 only the vector equations. Comparing with eq.(B.1),
 $$k=-{\kp\over 2},\;\; \hat k={\kp\over 2}\; .\eqno(B.14)$$
 Thus by eqs.(B.8)
$$\be_+=\be_+(u^+),\;\;\be_-=\be_-(u^-)\eqno(B.15)$$
 and similarly for $\hat\be_\pm$; (B.13) therefore simplifies to
$$\be_+(u^+)\be_-(u^-)=\hat\be_+(u^+)\hat\be_-(u^-)\; ,\eqno(B.16)$$
 which implies,
$$\hat\be_+(u^+)=r_1\be_+(u^+)\eqno(B.17a)$$
$$\hat\be_-(u^-)={1\over{r_1}}\be_-(u^-)\eqno(B.17b)$$
where $r_1= const$. From the definitions (B.5b),(B.6b) and the
 scalar equation of motion (22), we now see that in fact $r_1=1$;
 and thanks to eqs.(B.17), we see that eq.(22), despite being a pair
 of equations among
  {\it functions}, is actually only a single numerical condition,
 once the vector equations of motion are used. \par
The functions $\be_+(u^+),\be_-(u^-)$ can be chosen freely. Let us
 restrict our attention to a region on $S$ where $\be_\pm$ have
 fixed signs, $s_\pm$. Perform a conformal (orthonormality
preserving) coordinate transformation on the worldsheet, in order
 to set $\be_\pm$ to the constants $s_\pm$; the new coordinates
 $\bar u^\pm$ are defined by $$d\bar u^\pm=\sqrt{\vert\be_\pm\vert}
du^\pm\; ,\eqno(B.18)$$ and the new worldsheet conformal factor is
$$\bar E=E/\sqrt{\vert\be_+\be_-\vert}\; .\eqno(B.19)$$
 For notational convenience, we henceforth suppress the bar on all
quantities. It must be kept in mind, though, that the residual
 worldsheet gauge freedom has now been used up. In the new
  coordinates, eqs.(B.12) become
$$E\nab\ln(-E)=-s-{{\kp^2}\over 4}E^2\eqno(B.20)$$
where $s=-s_+s_-$ is another sign. Recalling that $E<0$, we define
the Liouville mode $\phi$ of the classical, induced worldsheet
metric, thus:
$$E\equiv -e^\phi\eqno(B.21)$$ and we obtain from (B.20) the
differential equation:
$$\nab\phi+se^{-\phi}={{\kp^2}\over 4}e^\phi\; .\eqno(B.22)$$
{\bf (III) \ Solving for $x(u)$,$\hat x(u)$ in terms of $E(u)$:}
As above, we consider a region of $S$ in which $\be_\pm$ have fixed
 signs. Since reasonable choices for these functions vanish only
 at discrete values of $u^\pm$ (respectively), the method presented
 here can reproduce any generic solution over most of the worldsheet.
 We will present the method for $x(u)$, but again, the same procedure
 can be applied to $\hat x(u)$ provided $\kp$ is replaced with
 $-\kp$. \par
 As in part (II), we choose local orthonormal coordinates $u^\pm$
 where $\be_\pm=s_\pm$. Eq.(25a) then holds, namely
$$\prtpm x\times\prtpm^2 x=\pm s_\pm\prtpm x\; . \eqno(B.23)$$
Defining the null three-vectors $a_\pm\equiv\pm s_\pm\prtpm x$,
we find $$a_\pm\times\prtpm a_\pm=a_\pm\eqno(B.24a)$$
        $$(a_\pm)^2=0\; .\eqno(B.24b)$$
Form the cross-product of eq.(B.24a) with $\prtpm a_\pm$ and
use (A.1):
  $$(\prtpm a_\pm)^2=1\eqno(B.24c)$$
 It is now a simple matter to solve eqs.(B.24). The general solution
is $$a^\mu_\pm={1\over{\prtpm\gm_\pm}}(1,\cos\gm_\pm,\sin\gm_
\pm)\eqno(B.25)$$
where $\gm_\pm$ are two unknown functions on the worldsheet.\par
 Let us now impose the integrability condition
$$\prtm a_+=s\prtp a_-\; ,\eqno(B.26a)$$
 which follows from the definition of $a_\pm$, and the vector
equation of motion, which reads\foot{The scalar equation (22) was
already used up in showing that $\be_\pm=\hat\be_\pm$ (see part
 (II)).}
$$\prtm a_+={\kp\over 2}s_-a_+\times a_-\; .\eqno(B.26b)$$
 Here $s=-s_-s_+$, as defined after eq.(B.20).  \par
After some algebra, we obtain from (B.25)-(B.26) the following useful
 equations:
$${\kp\over 2}s_-\ll 1-\cos(\gm_+-\gm_-)\rr=s(\prtp\gm_-)(
\prtp\gm_+)=-(\prtm\gm_-)(\prtm\gm_+)\; .\eqno(B.27)$$
On the other hand, the worldsheet conformal scale factor is, by
definition,
$$E=\prtp x\cdot\prtm x=sa_+\cdot a_-=-s{1\over{(\prtp\gm_+)
(\prtm\gm_-)}}\ll 1-\cos(\gm_+-\gm_-)\rr\eqno(B.28)$$
 Eqs.(B.27)-(B.28) give (25b); the rest of the procedure is
 described in section 4.
\medskip{\bf Appendix C:
\ Constancy of End Masses --- Geometric Proof.}\smallskip
We now prove that the geometrically defined endmasses for the
 open string (see section 3) are constants, that is, $\tau$
independent. The proof is for the $P$ end, but carries over
  to $Q$. \par
  The instantaneous mass $m$ is the angle of the wedge
 between the top and bottom worldsheets, at $P$ and in its
instantaneous rest frame (the `deficit angle'). Since the coordinates
 $(\tau,\sigma)$ are orthonormal and $\sigma=\; const$ at $P$, we
 have\foot{We denote $\vert a\vert\equiv\sqrt{-a^2}$ for a
 timelike vector $a$. The angle
 $m$ is positive, and bounded from above as
discussed in section 5. Equation (C.1) requires a sign correction if
$m>\pi$.}
$$\sin m={{\vert
{{\prt x}\over{\prt\sigma}}\times{{\prt\hat x}\over{\prt\sigma}}
\vert}\over{({{\prt x}\over{\prt\sigma}})^2}}={{\vert
{{\prt x}\over{\prt\sigma}}\times{{\prt\hat x}\over{\prt\sigma}}
\vert}\over{-({{\prt x}\over{\prt\tau}})^2}} \eqno(C.1)$$
 Thanks to orthonormality and the boundary condition, we have at $P$:
$$(\xot)^2=-(\xos)^2,\; (\hxot)^2=-(\hxos)^2 \eqno(C.2a)$$
$$\hxot=\xot\eqno(C.2b)$$
Thus we may define the following two unit spacelike vectors:
$$a\equiv{1\over{\vert\xot\vert}}\xos,\;
b\equiv{1\over{\vert\xot\vert}}\hxos,\;a^2=b^2=1\eqno(C.3)$$
All equations from here on will be understood to hold at $P$.
Clearly $(n\times\hat n)\vert\vert\xot$, and
$$n^2=\hat n^2=1,\; a\cdot n=b\cdot\hat n=0,\;(n\times\hat n)\cdot
a=(n\times\hat n)\cdot b=0\; .\eqno(C.4)$$
Next, invoke the scalar equations (22), from which follows
\foot{By subtracting the two versions with $\pm$ signs one from the
 other.}
$$\xot\times\xos\cdot\xost=\xot\times\hxos\cdot
\hxost\; ,\eqno(C.5)$$
and thus $$n\cdot{{\prt a}\over{\prt\tau}}=\hat n\cdot
{{\prt b}\over{\prt\tau}} \eqno(C.6)$$
 Since $a$,$b$ are unit vectors, $a\cdot\prt a/\prt\tau=b\cdot\prt b
/\prt\tau=0$; this, together with (C.4) and (C.6), allows us to expand
 (with $\al,\be,\gm,\delta$ unknown functions of $\tau$),
$${{\prt a}\over{\prt\tau}}=\al a\times n+\be(a\cdot\hat n)n
\eqno(C.7a)$$
$${{\prt b}\over{\prt\tau}}=\gm b\times\hat n+
\delta(b\cdot n)\hat n
\eqno(C.7b)$$
$$\be a\cdot\hat n=\delta b\cdot n\eqno(C.7c)$$
 We thus find:
 $${\prt\over{\prt\tau}}(a\cdot b)=\be(a\cdot\hat n)(b\cdot n+
 a\cdot\hat n)\eqno(C.8)$$
But $a,b,n,\hat n$ are all unit vectors in the two-dimensional
spatial subspace of the instantaneous rest-frame at $P$. Thus,
eq.(C.4) implies the vanishing of the right-hand side in eq.(C.8),
and (C.1) then gives
   $${{\prt}\over{\prt\tau}}m=0\; ,\eqno(C.9)$$
 since $a\cdot b=\cos m(\tau)$. Thus the endmass at $P$ is conserved,
as claimed.
\medskip{\bf Appendix D:
\ Flat Open String with End Masses.}\smallskip
In this appendix, we treat the three-dimensional
classical open string in the flat limit (no gravity), with masses
at the endpoints. We use non-geometric methods, and the results
agree with the flat limit of the geometric formalism, described in
section 6. We also describe the large endmass
 (or small string-tension) expansion for the flat case,
  and present as an
  example a simple infinite-mass configuration
 where the two endpoints pass each other with an impact parameter.
 \par
The flat equation of motion is the worldsheet wave equation,
 $$\nab x^\mu=0\; ,\eqno(D.1)$$
 with the general solution $$x^\mu=A^\mu(\tau+\sigma)+\tilde A^\mu
(\tau-\sigma)\; .\eqno(D.2a)$$
 As usual, we choose an orthonormal gauge on the worldsheet so that
 $P$,$Q$ are at $\sigma=0,\pi$, respectively; as pointed out in the
 footnote following eq.(6), this still leaves the freedom to
reparametrize $u^+\rightarrow f(u^+)$, $u^-\rightarrow f(u^-)$,
 where $f(v)-v$ is periodic with period $2\pi$. \par
 In such a parametrization, eqs.(30b) hold, namely
$$(A^\prm)^2=(\tilde A^\prm)^2=0. \eqno(D.2b)$$
  In addition, the total momentum
  of the bulk of the string (without
 the ends) is $\kp\int_0^\pi d\sigma{{\prt x^\mu}\over{\prt\tau}}$,
 and the total momentum of the system\foot{ An equivalent
 expression for the bulk string momentum is the equal-time
 spatial integral $\int d^2{\bf x} T^{0\mu}$;
  this expression is, of course, valid in any worldsheet
  parametrization, though the expression (11) is not.},
  including the ends, is:
 $$p_{total}^\mu=\kp\int_0^\pi d\sigma{{\prt x^\mu}\over
 {\prt\tau}}+\sum_{P,Q}m{{dx^\mu}\over{ds}}\; ,\eqno(D.3)$$
 where $s$ is proper time at either endpoint. As in the text, we
 assume for simplicity that the endmasses are equal, $m^{(P)}=
m^{(Q)}=m$.\par
 Imposing momentum conservation $dp_{total}/d\tau=0$, we
  integrate by parts and use eq.(D.1); this yields the
 boundary conditions
 $$\kp\xos(P)=m{d\over{d\tau}}({{dx(P)}\over{ds}})\eqno(D.4a)$$
 $$\kp\xos(Q)=-m{d\over{d\tau}}({{dx(Q)}\over{ds}})\eqno(D.4b)$$
 By eq.(D.2a), an element of proper time at $P$ is given by
 $$ds^2=-2A\upm(\tau)\cdot\tilde A\upm(\tau)d\tau^2\; ,\eqno(D.5)$$
with a similar relation holding at $Q$.\par
 We define (see eq.(34b)) $$\rho\equiv m/\kp\eqno(D.6)$$
and use eq.(D.5) to rewrite the boundary conditions (D.4):
$$A\upm(\tau)-\tilde A\upm(\tau)=\rho{d\over{d\tau}}\{
{1\over{\scriptstyle \sqrt{-2A\upm(\tau)\cdot\tilde A\upm(\tau)}
}}(A\upm(\tau)
+\tilde A\upm(\tau))\}, \eqno(D.7a)$$
$$A\upm(\tau+2\pi)-\tilde A\upm(\tau)=-\rho{d\over{d\tau}}\{
{1\over{\scriptstyle
\sqrt{-2A\upm(\tau+2\pi)
\cdot\tilde A\upm(\tau)}}}(A\upm(\tau+2\pi)
+\tilde A\upm(\tau))\}. \eqno(D.7b)$$
 When $\rho=0$, eqs.(D.7) imply that $A\upm=\tilde A\upm$ and that
 $A\upm$ is periodic with period $2\pi$, and the standard open
 string theory is recovered\foot{This limit should be taken carefully,
 since then $ds/d\tau=0$ (lightlike endpoints), so it is not
 immediately clear that the r.h.s. of eqs.(D.7) vanishes in the
 limit.}.    \par
 It suffices to examine in detail the boundary condition at $P$,
  since the
condition at $Q$ is obtained from it by the replacements
$$\rho\rightarrow-\rho,\; A(\tau)\rightarrow A(\tau+2\pi).
\eqno(D.8)$$
Eq.(D.7a) can be integrated, as both sides are $\tau$ derivatives;
 this introduces an unknown constant vector, $c^\mu$, and the
 boundary condition at $P$ assumes the form\foot{All arguments in
 eqs.(D.9) are $\tau$.}
$$\varphi(A\upm+\tilde A\upm)=c-{1\over\rho}(
\tilde A-A)\; ,\eqno(D.9a)$$
 where $$\varphi^{-2}\equiv-2A\upm\cdot\tilde A\upm\; .\eqno(D.9b)$$
 The l.h.s. of (D.9a) is $dx(P)/ds$, the three-velocity at $P$. Thus,
 in the limit $\rho\rightarrow\infty$ (zero string tension or
 infinite endmasses), $c^\mu$
              is the fixed velocity vector at $P$. \par
 Due to eq.(D.9b), the square of the l.h.s. of (D.9a) is $-1$;
 in fact, (D.9a)
  {\it alone} implies $\ll\varphi(A\upm+\tilde A\upm)\rr
\cdot(A\upm+\tilde A\upm)=0$, which in turn implies that
          $$\ll\varphi(A\upm+\tilde A\upm)\rr^2=\; const.$$

  To solve for the dynamics of the string (eq.(D.2a)),
   eqs.(D.9) and the corresponding $Q$ condition must be solved.
 For large $\rho$, this can be done by expansion in powers of
  $1/\rho$, as we now describe. We look for solutions for which
  $A$,$\tilde A$ are
 $O(\rho^0)$. Assuming $$A\upm\cdot c<0,$$ eqs.(D.9) give
$$\tilde A\upm=-A\upm-2(A\upm\cdot c)c+O(1/\rho)\; ,\eqno(D.10a)$$
$$c^2=-1+O(1/\rho)\eqno(D.10b)$$
((D.10b) follows from squaring eq.(D.9a) and using (D.9b)). Eq.(D.10a)
 ensures that $(\tilde A\upm)^2=0$ to lowest
  order in $1/\rho$, provided
 $(A\upm)^2$ vanishes to that order.\par
 By dotting (D.9a) with $A\upm$ and using (D.2b),(D.9b)
  and eqs.(D.10),
$\varphi^{-1}$ is found to order $1/\rho$. Then we use (D.9a)
 again, and find the\foot{$A$,$\tilde A$ may be shifted by arbitrary
 constant vectors, provided $c$ is adjusted so that (D.9a) still holds.
 Modulo this freedom, our $1/\rho$-perturbative solution is unique.}
  next-order solution for $\tilde A$ in terms of $A$:
 $$\tilde A\upm=-A\upm-2(A\upm\cdot c)c-{4\over\rho}(A\upm\cdot c)A
-{4\over\rho}c\ll A\upm\cdot A+2(A\cdot c)(A\upm\cdot c)\rr+O(
{1\over{\rho^2}})\; ,\eqno(D.11a)$$
$$c^2=-1+O({1\over{\rho^2}})\; ,\eqno(D.11b)$$ where the
 integration constant for $\tilde A$ is so far only determined
 to $O(\rho^0)$:
$$\tilde A=-A-2(A\cdot c)c+O(1/\rho)\; .\eqno(D.11c)$$
The $O(1/\rho)$ integration constant for $\tilde A$ can be determined,
 once we choose $c^2$ to order $O(1/\rho^2)$. It is easiest to
 simply choose $c^2=-1$ to all orders. \par
 $A(\tau)$ is so far unconstrained, except for $(A\upm)^2=0$ and the
 assumed
 $$A\upm\cdot c<0\; .\eqno(D.12)$$
 Due to the former, and to $c^2<0$, it is sufficient to
 assume $A^{\prm 0}>0$ to ensure eq.(D.12), since\foot{Both $c^0>0$
 and $c^2<0$ follow, for $\rho$ sufficiently large, from eq.(D.9a),
 since $dx/ds$ is timelike and $dx^0/ds>0$.}
  $c^0>0$.\par
   The boundary condition at $Q$ is similarly solved, order by
 order; this gives {\it another} expression for $\tilde A(\tau)$,
 in terms of $A(\tau+2\pi)$, $-1/\rho$ and $d^\mu$ (the constant
 vector arising in the integration of eq.(D.7b)).
 Comparison of the two expressions yields a vector relation between
 $A(\tau)$ and $A(\tau+2\pi)$; this relation modifies the periodicity
 of $A\upm$, and therefore of $\tilde A\upm$. \par
 To get a flavor for how this works, we restrict attention to the
 $\rho\rightarrow\infty$ limit itself. Physically, this corresponds
 either to two free masses with no string (an uninteresting case in
 the absence of gravity!), or two infinitely massive particles, with
 a string between them, moving past each other. \par
 In this limit, eq.(D.11c) becomes
 $$\tilde A(\tau)=-A(\tau)-2(A(\tau)\cdot c)c\; ,\eqno(D.13a)$$
 and the corresponding boundary condition at $Q$ is,
 $$\tilde A(\tau)=-A(\tau+2\pi)-2(A(\tau+2\pi)\cdot d)d\; .
 \eqno(D.13b)$$
 Eliminating $\tilde A$, we find
$$A\upm(\tau)+2(A\upm(\tau)\cdot c)c=A\upm(\tau+2\pi)+2(A
\upm(\tau+2\pi)\cdot d)d\eqno(D.14a)$$
But the operation $v\rightarrow v+2c(c\cdot v)$ on vectors $v$ is
 a reflection, and so is the corresponding operation with $d$.
 Thus, solving (D.14a) for $A\upm(\tau+2\pi)$ gives
$$A\upm(\tau+2\pi)=\Lambda^{(0)}\cdot A\upm(\tau)\eqno(D.14b)$$
where $\Lambda^{(0)}$ is a Lorentz transformation.
 In the center-of-mass
 frame of the two endmasses, let us choose the space axes such that
 $$c^\mu=(\cosh\chi,0,\sinh\chi),\; d^\mu=(\cosh\chi,0,-\sinh\chi)\;
.\eqno(D.15)$$
 Let us further denote

 $$\Lambda(\om)^\mu_{\;\;\nu}\equiv
\left( \eqalign{\cosh\om\quad\quad
&0\quad -\sinh\om\quad\cr
  0\quad\quad&1\quad\quad 0\cr
         -\sinh\om\quad\;&0\quad\;\cosh\om\cr}\right)
 \eqno(D.16a)$$
 Then, from eqs.(D.14),
  $$\Lambda^{(0)}=\Lambda(4\chi)\; .\eqno(D.16b)$$
 The physical interpretation of the modified periodicity (D.14b) is
 simple: an endpoint, being infinitely massive, totally reflects
 string waves in the rest frame of the mass. But the ends move
  with
 rapidities $\pm\chi$, so in the center-of-mass the waves undergoe
 a Doppler shift upon each reflection, by a boost of rapidity
 $2\chi$. Thus, by the time a wave reflects once from each boundary,
 it has undergone a boost of rapidity $4\chi$. \par
  The general solution of the functional equation $(D.14b)$ is:
 $$A\upm(\tau)=\Lambda(2\chi\tau/\pi)\cdot D(\tau)\; ,\eqno(D.17)$$
 where $D^\mu(\tau)$ is any lightlike vector which is a periodic
 function of $\tau$ (with period $2\pi$). To guarantee eq.(D.12),
 it suffices to impose $$D^0>0.$$ \par
 Once $D$ is chosen, we have a leading-order solution, and eqs.(D.11)
  (with their couterparts at $Q$) can be used to find a functional
 equation for the $O(1/\rho)$ piece of $A(\tau)$; this can again be
 solved by means of periodic functions and hyperbolic functions
 of $2\chi\tau/\pi$, and so on to any desired order in the
  $O(1/\rho)$ expansion.\par Returning to order $O(\rho^0)$,
  we conclude with a simple example of a string configuration of the
 type (D.17). Make the following simple choice for $D(\tau)$,
$$D^\mu(\tau)=(1,1,0)\; .\eqno(D.18)$$
 We then find from eqs.(D.13) and (D.17), after making
 an arbitrary choice for $A(0)$, that
 $$A^\mu(\tau)=({1\over\al}\sinh\al\tau,\tau,
 -{1\over\al}\cosh\al\tau)\; ,\eqno(D.19a)$$
 $$\tilde A^\mu(\tau)=({1\over\al}\sinh\al(\tau+\pi),-\tau,
 {1\over\al}\cosh\al(\tau+\pi))\; ,\eqno(D.19b)$$
where $$\al\equiv 2\chi/\pi\; .\eqno(D.19c)$$
 Substituting eqs.(D.19) in (D.2a), and denoting $x^\mu=(t,x,y)$,
 we obtain the string configuration in the COM frame:
 $$y=-t\tanh\ll{\al\over 2}(x-\pi)\rr \eqno(D.20a)$$
with the ranges $$-\infty<t<\infty,\; 0\leq x\leq 2\pi \eqno(D.20b)$$
 where the end $P$ ($Q$) is at $x=0$ ($x=2\pi$). In this configuration
 (see Fig. 4), the instantaneous shape of the string is a section of
 a hyperbolic-tangent curve, symmetric about the string center
  $(x,y)=(\pi,0)$; the section
 becomes the complete curve in the limit of infinite endpoint
 rapidities, and the relative scale of the $x$,$y$ coordinates
  of the curve is linear in
  Minkowski time. The masses move in opposite
 directions, parallel to the $y$ axis, and their separation at
 closest approach is along the $x$ axis and equal\foot{
 This separation can be scaled to any desired number, by rescaling
 the choice for $D^\mu$ (eq.(D.18)) by an arbitrary positive
  number.} to $2\pi$.
\ACK{
 I would like to thank Lenny Susskind, J. Bjorken, Larus Thorlacius
 and Adrian Cooper for discussions on closed timelike curves
 and their generation at finite time, and on particle kinematics in
 $2+1$ dimensions.
 }
\endpage
\refout
\endpage
{\bf\ \ \ \ \ \ \ \ \ FIGURE CAPTIONS}\bigskip
{\bf Fig. 1:}\ \ \ A local equal-time section of the worldsheet.
                   Endpoints $P$,$Q$ are connected by the string.
\medskip
{\bf Fig. 2:}\ \ \ A section of the $\{z\}$ coordinate patch. $S$ is
                   the worldsheet and $T$ the transition surface
                    (each has a top and bottom side). Arrows indicate
                    indentifications.
\medskip
{\bf Figs. 3:}\ \ \ Fig. 2 for two free, static masses;
       (a) Spatial section open.
                      (b) Spatial section closes due to excessive
                     masses.
\medskip
{\bf Fig. 4:}\ \ \ The example flat-string configuration of Appendix
                     D.
 \endpage
\bye